\documentclass[letterpaper]{article} 

\ifdefined\aaaianonymous
    \usepackage[submission]{aaai2026}  
\else
    \usepackage{aaai2026}              
\fi

\usepackage{tabularx}
\usepackage{booktabs}
\usepackage{amsmath}
\usepackage{amssymb}
\usepackage{multirow}
\usepackage{pifont}
\usepackage{times}  
\usepackage{helvet}  
\usepackage{courier}  
\usepackage[hyphens]{url}  
\usepackage{graphicx} 
\usepackage{natbib}  
\usepackage{caption} 
\usepackage{algorithm}
\usepackage{algorithmic}

\usepackage{newfloat}
\usepackage{listings}
\DeclareCaptionStyle{ruled}{labelfont=normalfont,labelsep=colon,strut=off} 
\floatstyle{ruled}
\newfloat{listing}{tb}{lst}{}
\floatname{listing}{Listing}

\title{Simulating Dispute Mediation with LLM-Based Agents for Legal Research}

\author{
    Junjie Chen\textsuperscript{\rm 1},
    Haitao Li\textsuperscript{\rm 1},
    Minghao Qin\textsuperscript{\rm 2},
    Yujia Zhou\textsuperscript{\rm 1},
    Yanxue Ren\textsuperscript{\rm 3},
    Wuyue Wang\textsuperscript{\rm 4},
    Yiqun Liu\textsuperscript{\rm 1},
    Yueyue Wu\textsuperscript{\rm 1}\thanks{Corresponding Author: wuyueyue1600@gmail.com},
    Qingyao Ai\textsuperscript{\rm 1}\thanks{Corresponding Author: aiqy@tsinghua.edu.cn}
}
\affiliations{
    \textsuperscript{\rm 1}Department of Computer Science and Technology, Tsinghua University
    \textsuperscript{\rm 2}China University of Political Science and Law
    \textsuperscript{\rm 3}Artificial Intelligence Laboratory, Bosera Asset Management Co., Ltd.
    \textsuperscript{\rm 4}University of Notre Dame
}

\usepackage{bibentry}

\begin{document}

\maketitle

\begin{abstract}
Legal dispute mediation plays a crucial role in resolving civil disputes, yet its empirical study is limited by privacy constraints and complex multivariate interactions. To address this limitation, we present \textit{AgentMediation}, the first LLM-based agent framework for simulating dispute mediation. It simulates realistic mediation processes grounded in real-world disputes and enables controlled experimentation on key variables such as disputant strategies, dispute causes, and mediator expertise.
Our empirical analysis reveals patterns consistent with sociological theories, including Group Polarization and Surface-level Consensus. As a comprehensive and extensible platform, \textit{AgentMediation} paves the way for deeper integration of social science and AI in legal research. 
\end{abstract}

\begin{links}
    \link{Code}{https://github.com/cjj826/AgentMediation}
\end{links}


\section{Introduction}
Legal dispute mediation, as a key form of alternative dispute resolution (ADR)~\cite{mnookin1998alternative}, plays an important role in the resolution of civil disputes worldwide. Compared to litigation, mediation is generally less time-consuming and more cost-effective, offering a promising approach to reducing court caseloads and accelerating the dispute resolution process~\cite{sherman2025alternative}. 

Much research in law and sociology has tried to investigate factors affecting the effectiveness of dispute mediation, including the context of the dispute, the procedural design of the mediation process, the behavioral strategies of the parties to the dispute (disputants), and the domain expertise of mediators~\cite{hsieh2022predicting}. However, systematically exploring dispute mediation remains challenging due to two primary factors: (1) privacy constraints, i.e., most mediation processes are confidential and not publicly accessible, making it difficult to collect large-scale real-world data for statistical analysis or experimental validation; and (2) multivariate interactions, i.e., dispute mediation outcomes are jointly affected by multiple variables, which are hard to isolate and quantify their individual effects in real-world settings.

\begin{figure*}[t!]
    \centering
    \includegraphics[width=0.9\textwidth]{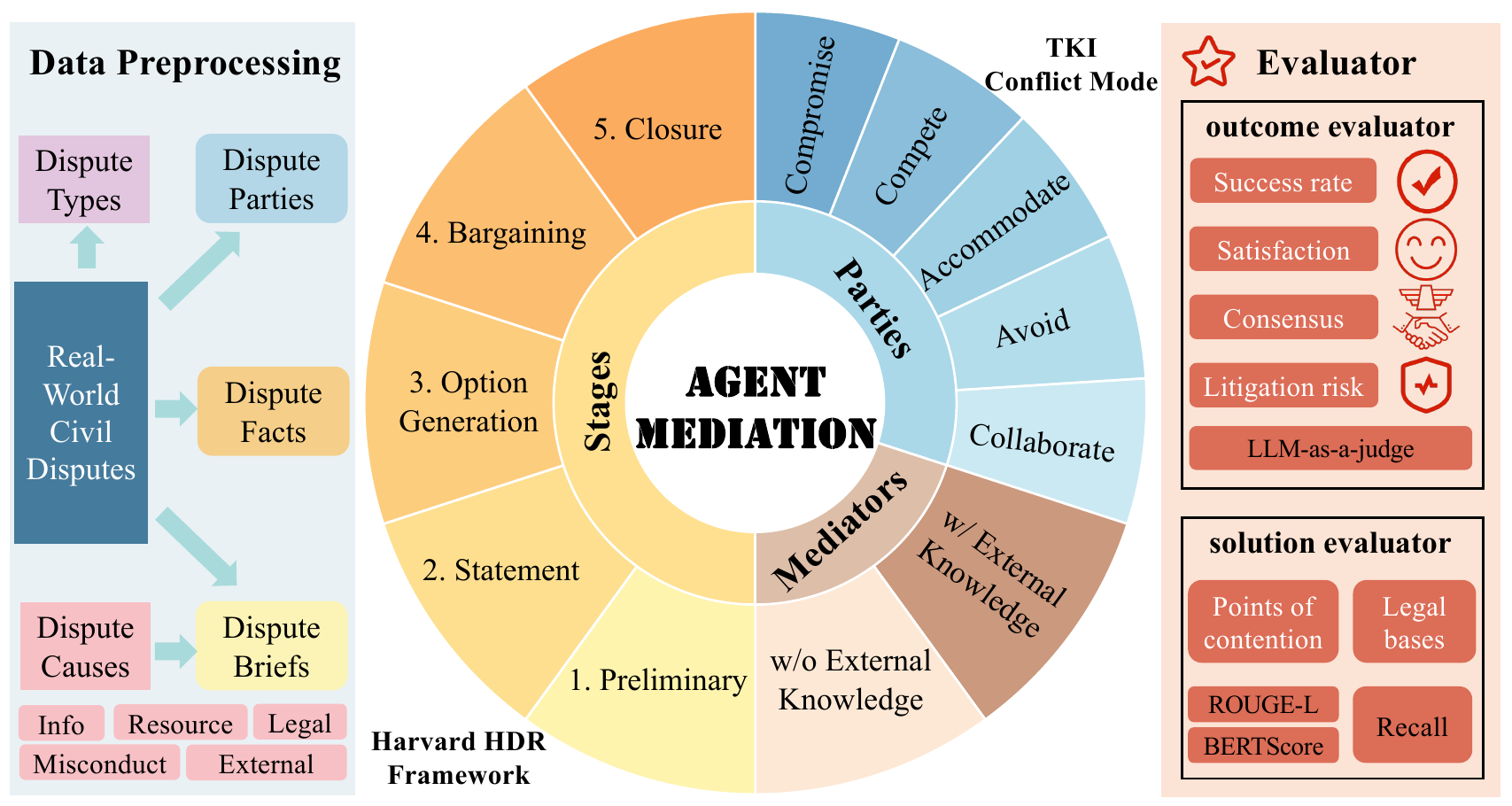}
    \caption{Overview of the \textit{AgentMediation} framework: (1) data preprocessing from real-world civil disputes to extract structured inputs for simulation; (2) a five-stage mediation process inspired by the Harvard HDR framework; (3) configurable agent roles, including parties with TKI-based conflict modes~\cite{thomas2008thomas} and mediators with or without access to external knowledge; and (4) a dual evaluation system for assessing both mediation outcomes and solution quality.}
    \label{fig:enter-label} 
\end{figure*}

Recent advances in LLM-based agent simulations~\cite{park2023generative,ashery2025emergent} have shown that multi-agent systems can serve as controllable and observable social laboratories for modeling complex human interactions. More specifically, studies like AgentsCourt~\cite{he2024agentscourt} demonstrate the feasibility of simulating legal scenarios with LLM-based agents. However, simulating dispute mediation with LLM-based agents remains a non-trivial task because of three reasons: (1) the scarcity of high-quality and publicly available mediation data, which limits the realistic and effective simulations; (2) the absence of a comprehensive and structured framework to model key factors in mediation and analyze their impact on outcomes; and (3) the lack of reliable and efficient quantitative evaluation methods for assessing mediation performance.

To this end, we introduce \textit{AgentMediation} (Figure \ref{fig:enter-label}), the first LLM-based agent framework for simulating dispute mediation that offers a controllable, reproducible and extensible platform. Specifically, this consists of three core components: data preprocessing, mediation simulation framework, and evaluator. First, to establish a reliable and open data foundation, we processed 330 civil disputes from the \textit{Dispute Resolution Case Database}\footnote{open at \url{https://dyjfalk.court.gov.cn/site}}, an authentic yet semi-structured dataset released by the Supreme People’s Court of China after ten months of collection, with the mediation process omitted to preserve privacy. We applied automatic extraction followed by manual refinement to construct structured representations that support realistic mediation simulation and evaluation. Second,  building on the above data, \textit{AgentMediation} provides a unified and controllable framework that follows a general five-stage mediation process inspired by the Harvard handbook of dispute resolution (HDR)~\cite{moffitt2012handbook}. We model three core aspects of mediation dynamics based on the HDR framework, namely the disputant behavior, dispute causes, and mediator expertise. Third, to support reliable and efficient evaluation, we design a dual-perspective assessment framework that captures mediation outcomes and solution quality. From the perspective of dispute mediation outcome, we adopt metrics such as mediation success rate, participant satisfaction, consensus level, and litigation risk. From the perspective of dispute mediation solution, we assess how well the mediators identify the points of contention and provide sound legal bases when necessary.

Our contributions fall into three aspects:

\noindent\textbf{1. Comprehensive Framework}: We build \textit{AgentMediation}, the first framework that uses LLM-based agents to simulate full dispute mediation processes, enabling fine-grained and dynamic analysis of how various factors affect mediation outcomes. Through carefully designed human evaluations and annotations, we show the effectiveness and reliability of our system in simulating real dispute mediation processes.
 
\noindent\textbf{2. Large-Scale and Extensible Dataset}: Based on 330 real civil disputes, we create a large-scale dataset with over 14,000 simulated mediation processes, including detailed dialogues and solutions. Our framework further supports on demand generation of diverse scenarios for future research.

\noindent\textbf{3. Theory-Aligned Empirical Insights}: Using the \textit{AgentMediation}, we find results that match social science theories, including Group Polarization~\cite{moscovici1969group}, Surface-level Consensus~\cite{kelman1958compliance,habermas1985theory}, Moral Emotions Theory~\cite{haidt2003moral}, Realistic Conflict Theory~\cite{Sherif1961RobbersCave}, and the Common In-Group Identity Model~\cite{gaertner1993common}. These insights demonstrate the potential of LLM-based agent simulation to capture legal-social dynamics, contributing to the development of future intelligent mediation systems.


\section{Related Work}
\begin{table*}[t]
\small
\setlength{\tabcolsep}{0.6mm} 
\centering
\begin{tabular}{lccccc}
\toprule
\textbf{Feature}                & \begin{tabular}[c]{@{}c@{}}GENTEEL NEGOTIATOR\\ \cite{priya2025genteel}\end{tabular} & \begin{tabular}[c]{@{}c@{}}LLM-Stakeholders \\ \cite{abdelnabi2024cooperation}\end{tabular} & \begin{tabular}[c]{@{}c@{}}WarAgent \\ \cite{hua2023war}\end{tabular}      & \begin{tabular}[c]{@{}c@{}}AgentsCourt \\ \cite{he2024agentscourt}\end{tabular}   & \begin{tabular}[c]{@{}c@{}}AgentMediation \\ (Ours)\end{tabular} \\ \midrule
Negotiation Dialogue Generation & \ding{52} & \ding{52} & \ding{52} & \ding{52} & \ding{52} \\
Multi-Agent Simulation          & \ding{56}     & \ding{52} & \ding{52} & \ding{52} & \ding{52} \\
Real-World Civil Disputes      & \ding{56}     & \ding{56}     & \ding{56}     & \ding{56}     & \ding{52} \\
Mediation Simulation Framework & \ding{56} & \ding{56} & \ding{56} & \ding{56} & \ding{52} \\
Mediation Evaluation Metrics & \ding{56} & \ding{56} & \ding{56} & \ding{56} & \ding{52} \\
\bottomrule
\end{tabular}
\caption{Comparison of existing LLM-based agent simulation works. Current approaches are limited in modeling the dynamics of dispute mediation, which our approach is designed to address.}
\label{tab:simulation-comparison}
\end{table*}

\textbf{Research on Legal Dispute Mediation.} Dispute mediation plays a crucial role in resolving civil disputes efficiently and reducing judicial burdens~\cite{mnookin1998alternative, sherman2025alternative}. Prior studies in law, sociology, and computational methods have tried to examine how factors such as dispute context, procedural design, party behavior, and mediator expertise affect mediation outcomes~\cite{hsieh2022predicting}. However, existing research often relies on datasets that are either private or limited in scale~\cite{tan2024robots}, restricting reproducibility and broader empirical analysis. In addition, these datasets are typically static and non-scalable~\cite{chawla2021casino, hale2025kodis}, making it difficult to capture and disentangle the effects of complex, interacting variables. As a result, studying the dynamic nature of mediation processes remains a significant challenge.

\textbf{LLM-Based Agents.} Recent work has shown that LLMs can serve as autonomous agents capable of reasoning, role-playing, and interacting within multi-agent settings. Studies such as Generative Agents~\cite{park2023generative} simulates social dynamics in virtual towns, while AgentsCourt~\cite{he2024agentscourt} models courtroom debates. These advances~\cite{jin2024agentreview, wu2023autogen} reveal LLMs' potential to reproduce complex human behaviors in controlled, repeatable settings. However, as Table \ref{tab:simulation-comparison} shows, using LLM-based agents to simulate dispute mediation remains highly challenging, primarily due to the scarcity of real-world data, the lack of frameworks tailored to mediation, and the absence of reliable and efficient quantitative evaluation methods.

\section{Methodology}
As Figure \ref{fig:enter-label} shows, \textit{AgentMediation} is designed as an extensible testbed to study the impact of various factors on mediation outcomes.  In this section, we introduce the details of \textit{AgentMediation}, focusing on five core components: Data Preprocessing, Mediation Process, Roles, Dispute Causes Taxonomy, and Evaluator.

\subsection{Data Preprocessing}
As previously discussed, we use the \textit{Dispute Resolution Case Database} as the foundation of our study. We denote this data resource as $\mathcal{D} = \{D_1, D_2, \dots, D_N\}$, where each sample $D_i$ represents a civil dispute. Each dispute $D_i$ follows a fixed schema and can be formalized as:
\begin{equation}
D_i = (\text{Title}_i, \text{Keywords}_i, \text{Brief}_i, \text{Method}_i, \text{Bases}_i),
\end{equation}
where $\text{Title}_i$ is the case title, $\text{Keywords}_i$ is a set of descriptive tags, $\text{Brief}_i$ provides the case background, $\text{Method}_i$ outlines the resolution approach (excluding detailed mediation dialogues), and $\text{Bases}_i$ contains the corresponding legal bases (i.e., statutory provisions) annotated by human experts.

To prepare the data for simulation, we use GPT-4o (2024-08-06)~\cite{achiam2023gpt} to extract key components, followed by manual verification to ensure quality. We first extract the dispute type ($DT_i$) from the $\text{Keywords}_i$. Then the $\text{Brief}_i$ is used directly as the dispute brief ($DB_i$) for our mediation simulation. From the $\text{Method}_i$, we extract objective factual statements as dispute facts ($DF_i$), providing supporting evidence for the mediator. Simultaneously, we identify the dispute parties ($DP_i$) from the $\text{Brief}_i$ for role-based interaction modeling. In addition, we extract the points of contention ($DPoints_i$) from $\text{Brief}_i$ and $\text{Method}_i$, and normalize the $\text{Bases}_i$ into a structured form ($DBases_i$). The overall data preprocessing can be summarized as:

\begin{align}
Di \xrightarrow{\text{Preprocess}} \big(& DT_i,\ DB_i,\ DF_i, \ DP_i,\nonumber \\
                                        & DPoints_i,\ DBases_i \big).
\end{align}

\begin{table}[t]
\centering
\small
\begin{tabular}{l r | l r}
\hline
\textbf{Attribute}         & \textbf{Value} & \textbf{Attribute}           & \textbf{Value} \\
\hline
\# of $D$                       & 330.00            & Avg. $DB$ Len.          & 160.0     \\
\# of $DT$        & 76.00             & Avg. \# of $DF$                      & 4.72     \\
Min \# of $DP$ & 2.00              & Avg. \# of $DBases$                  & 3.42   \\
Max \# of $DP$ & 6.00              &  Avg. $DPoints$ Len.     & 81.35  \\
\hline
\end{tabular}
\caption{Summary statistics of our dataset. ``\#'' denotes ``number''; ``Len.'' denotes length in Chinese characters.}
\label{tab:preprocess}
\end{table}

Table~\ref{tab:preprocess} presents the detailed statistics of the final preprocessed dataset. It is worth noting that \textit{AgentMediation} also supports user-defined disputes, allowing users to conveniently simulate customized mediation scenarios based on their specific needs. A concrete example can be found in Appendix Section 1.

\subsection{Mediation Process}
Inspired by the Harvard HDR framework~\cite{moffitt2012handbook}, \textit{AgentMediation} adopts a structured five-stage pipeline to simulate the mediation process: 

\noindent\textbf{I. Preliminary.} The system presents the dispute brief ($DB$) to help the mediators and parties build a shared understanding of the dispute. 

\noindent\textbf{II. Statement.} All parties and the mediators introduce themselves and state their positions and claims, preparing for further discussion. 

\noindent\textbf{III. Option Generation.} The mediators need to gather information and verify key facts. In our simulation, the preprocessed dispute facts ($DF$) are provided to simulate the outcome of the mediator’s fact-finding process (our system also supports user-defined dispute facts). Based on $DF$, the mediators are required to generate a preliminary proposal that includes the points of contention ($MDPoints$), supporting legal bases ($MDBases$), and a solution. 

\noindent\textbf{IV. Bargaining.} The system simulates multi-turn interactions where parties and mediators discuss the proposed resolution. Participants may agree, disagree, or propose alternatives, mimicking real-world negotiation dynamics. 

\noindent\textbf{V. Closure.} The bargaining stage ends when one of the following conditions is met: an agreement is reached, a disagreement is confirmed, or the round limit is reached. The mediator is then required to output a final solution.


\subsection{Roles}
In our \textit{AgentMediation} framework, we model two key roles: dispute parties (disputants) and mediators.

\noindent\textbf{Parties}. Parties serve as the primary roles for expressing disputes in the mediation process. Each party represents a distinct position, which may correspond to an individual, a group, or an institution. Even under the same dispute context, parties adopting different behavioral strategies can lead to significantly different mediation outcomes. To systematically investigate the impact of behavioral strategies on mediation performance, we follow the Thomas-Kilmann Conflict Mode Instrument (TKI)~\cite{thomas2008thomas}, which characterizes individual behaviors in dispute scenarios along two dimensions: assertiveness (the degree to which one seeks to satisfy their own concerns, $\alpha$) and cooperativeness (the degree to which one attempts to satisfy others’ concerns, $\beta$). Based on different levels of $\alpha$ and $\beta$ (High: H, Moderate: M, Low: L), five typical behavioral strategies are defined:\\
(1) \textbf{Competing} (H $\alpha$, L $\beta$): prioritizing self-interest and often dominating others;\\
(2) \textbf{Collaborating} (H $\alpha$, H $\beta$): seeking win-win outcomes through deep engagement;\\
(3) \textbf{Compromising} (M $\alpha$, M $\beta$): aiming for middle-ground solutions through mutual concession;\\
(4) \textbf{Avoiding} (L $\alpha$, L $\beta$): withdrawing from conflict;\\
(5) \textbf{Accommodating} (L $\alpha$, H $\beta$): placing others’ interests ahead of one’s own.

\textbf{Mediators.} Mediators play a central role in guiding the mediation process, facilitating communication, and proposing resolutions. Among the various factors that influence mediation quality, domain expertise is particularly crucial, especially for accurately interpreting and citing relevant legal bases. To investigate the impact of domain expertise, we simulate mediators with varying levels of domain expertise by controlling their access to an external legal knowledge source. In the \textit{without External Knowledge (w/o EK)} condition, mediators rely solely on the case context and their internal knowledge to cite legal bases. In the \textit{w/ EK} condition, mediators have access to a retrieval module\footnote{we use the API: \url{https://legalone.com.cn/}} that takes case descriptions as input and returns relevant legal bases.

\subsection{Dispute Causes Taxonomy}
\label{section: dct}
To address the limitations of traditional dispute type ($DT$) labels (e.g., housing contract), which typically reflect only surface-level dispute categories and fail to support the fine-grained causal reasoning required in mediation processes, we construct a dispute causes taxonomy. Specifically, we first design an initial framework based on existing research~\cite{vilendrer2023conflict}. Building on this framework, we leverage GPT-4o to classify the underlying causes of 1,080 real-world civil court cases\footnote{from \url{https://wenshu.court.gov.cn/}} based on their factual descriptions. When the existing taxonomy fails to accommodate certain cases, we dynamically extend the category set. The resulting taxonomy is then reviewed and refined by legal experts to ensure its completeness and interpretability. The final taxonomy consists of five top-level categories and twenty-nine subcategories. The top-level categories include: Information Conflict, Resource Conflict, Behavioral Misconduct, Legal Issues, and External Factors (e.g., natural disasters) (see Appendix Section 2 for details).

Importantly, this is not a single-label classification task but rather a multi-causal attribution process: a single case often involves multiple interrelated causes. Based on this taxonomy, we further investigate how different dispute causes impact the effectiveness of mediation.

\begin{table*}[t]
\small
\setlength{\tabcolsep}{1mm} 
\begin{tabular}{lcc|cc|cc|cc}
\hline
\multicolumn{1}{c}{\multirow{2}{*}{\textbf{\begin{tabular}[c]{@{}c@{}}Different \\ Backbone LLM\end{tabular}}}} & \multicolumn{2}{c|}{\textbf{Human1}}      & \multicolumn{2}{c|}{\textbf{Human2}} & \multicolumn{2}{c|}{\textbf{Human3}}      & \multicolumn{2}{c}{\textbf{Max of Three}} \\
\multicolumn{1}{c}{}                                                                                            & \textit{BERTScore (F1)}       & \textit{LLMScore}            & \textit{BERTScore (F1)}  & \textit{LLMScore}            & \textit{BERTScore (F1)}       & \textit{LLMScore}            & \textit{BERTScore (F1)}       & \textit{LLMScore}            \\ \hline
Llama-3.2-1B-Instruct                                                                                           & 0.6471$^{\dag\dag}$ & 0.4880$^{\dag\dag}$ & 0.6152         & 0.4530$^{\dag\dag}$ & 0.6424$^{\dag\dag}$ & 0.4780$^{\dag\dag}$ & 0.6629$^{\dag\dag}$ & 0.5310$^{\dag\dag}$ \\
Qwen3-0.6B                                                                                                      & 0.6647$^{\dag\dag}$ & 0.6280$^{\dag\dag}$ & 0.6353         & 0.5840$^{\dag\dag}$ & 0.6612$^{\dag\dag}$ & 0.6000$^{\dag\dag}$ & 0.6871$^{\dag\dag}$ & 0.6820$^{\dag\dag}$ \\
GLM-4-Flash                                                                                                     & 0.6857              & 0.7140$^{\dag\dag}$ & 0.6581         & 0.6580              & 0.6842              & 0.6440$^{\dag\dag}$ & 0.7141              & 0.7620              \\
GPT-4o-2024-11-20                                                                                               & 0.6954              & 0.7500              & 0.6553         & 0.6900              & 0.6895              & 0.6960              & 0.7200              & 0.8000              \\
DeepSeek-V3-0324                                                                                                & 0.6954              & 0.7780              & 0.6532         & 0.6780              & 0.6832              & 0.7020              & 0.7151              & 0.7940              \\ \hline
\end{tabular}
\caption{Human-model similarity scores across backbone LLMs. ${\dag}$ and ${\dag\dag}$ indicate statistically significant differences from DeepSeek-V3-0324 under a paired sample \textit{t}-test with $p < 0.05$ and $p < 0.01$, respectively.}
\label{realism-exp}
\end{table*}

\begin{table}[t]
\small
\setlength{\tabcolsep}{1mm} 
\begin{tabular}{llcccc}
\hline
\textbf{Stage}           & \textbf{Role}             & \textbf{Aspect}         & \multicolumn{1}{c}{\textbf{H}} & \multicolumn{1}{c}{\textbf{S}} & \multicolumn{1}{c}{\textbf{T}} \\ \hline
\multirow{7}{*}{Begin}   & \multirow{5}{*}{Mediator} & Fact Summarization      & 0.05                           & 0.10                           & \textbf{0.85}                  \\
                         &                           & Neutrality Statement    & 0.00                           & 0.25                           & \textbf{0.75}                  \\
                         &                           & Self Introduction       & 0.00                           & 0.00                           & \textbf{1.00}                  \\
                         &                           & Turn-Taking Guidance    & 0.00                           & 0.00                           & \textbf{1.00}                  \\
                         &                           & Preliminary Solution    & 0.10                           & \textbf{0.50}                  & 0.40                           \\ \cline{2-6} 
                         & \multirow{2}{*}{Parties}    & Claim Clarity           & 0.10                           & 0.00                           & \textbf{0.90}                  \\
                         &                           & Fact-Based Claims       & 0.05                           & 0.05                           & \textbf{0.90}                  \\ \hline
\multirow{8}{*}{During} & \multirow{4}{*}{Mediator} & Presiding Norms         & 0.05                           & 0.05                           & \textbf{0.90}                  \\
                         &                           & Debate Direction        & 0.00                           & 0.10                           & \textbf{0.90}                  \\
                         &                           & Order Control           & 0.00                           & 0.10                           & \textbf{0.90}                  \\
                         &                           & Fairness Awareness      & 0.25                           & 0.10                           & \textbf{0.65}                  \\ \cline{2-6} 
                         & \multirow{4}{*}{Parties}    & Order Compliance        & 0.00                           & 0.05                           & \textbf{0.95}                  \\
                         &                           & Offense Prohibition     & 0.05                           & 0.30                           & \textbf{0.65}                  \\
                         &                           & Effective Communication & 0.05                           & \textbf{0.50}                  & 0.45                           \\
                         &                           & Focused Debate          & 0.10                           & 0.05                           & \textbf{0.85}                  \\ \hline
\multirow{3}{*}{End}     & \multirow{3}{*}{Mediator} & Solution Feasibility    & 0.10                           & 0.00                           & \textbf{0.90}                  \\
                         &                           & Solution Legality       & 0.05                           & 0.00                           & \textbf{0.95}                  \\
                         &                           & Claim Coverage          & 0.10                           & 0.05                           & \textbf{0.85}                  \\ \hline
\end{tabular}
\caption{Legal experts’ pairwise preferences on 18 aspects across 20 disputes (H=Human, S=Simulation, T=Tie).}
\label{realism-exp-2}
\end{table}

\begin{table*}[t]
\small
\centering
\setlength{\tabcolsep}{1mm} 
\begin{tabular}{l|cccc|cccc}
\hline
\multirow{2}{*}{\textbf{Setting}} & \multicolumn{4}{c|}{\textbf{NUM = 1}}                     & \multicolumn{4}{c}{\textbf{NUM = ALL}}                      \\ \cline{2-9}
                                  & \textbf{\begin{tabular}[c]{@{}c@{}}Success Rate\\ ($SR\uparrow$)\end{tabular}} 
                                  & \textbf{\begin{tabular}[c]{@{}c@{}}Satisfaction\\ ($Sat\uparrow$)\end{tabular}} 
                                  & \textbf{\begin{tabular}[c]{@{}c@{}}Consensus\\ ($Con\uparrow$)\end{tabular}} 
                                  & \textbf{\begin{tabular}[c]{@{}c@{}}Litigation Risk\\ ($LR\downarrow$)\end{tabular}} 
                                  & \textbf{\begin{tabular}[c]{@{}c@{}}Success Rate\\ ($SR\uparrow$)\end{tabular}} 
                                  & \textbf{\begin{tabular}[c]{@{}c@{}}Satisfaction\\ ($Sat\uparrow$)\end{tabular}} 
                                  & \textbf{\begin{tabular}[c]{@{}c@{}}Consensus\\ ($Con\uparrow$)\end{tabular}} 
                                  & \textbf{\begin{tabular}[c]{@{}c@{}}Litigation Risk\\ ($LR\downarrow$)\end{tabular}} \\ \hline
Default Setting                          & 82\%                  & 54.77                 & 70.96               & 24.00                     & 82\%                  & 54.77                 & 70.96               & 24.00              \\ \hline
+ Compromising                      & 95\%                  & 55.72                 & 72.50               & 22.50                     & 99\%                  & 58.67                  & 75.25     & 19.25           \\
+ Competing                         & 29\%                  & 43.20$^{\dag\dag}$    & 53.25$^{\dag\dag}$  & 40.50$^{\dag\dag}$        & 11\%                  & 35.00$^{\dag\dag}$     & 33.00$^{\dag\dag}$ & 64.00$^{\dag\dag}$ \\
+ Accommodating                     & 88\%                  & 54.21$^{\dag}$        & 71.50               & 21.00                     & 97\%                  & 53.35$^{\dag\dag}$     & 75.50     & 16.75           \\
+ Avoiding                          & 54\%                  & 49.63$^{\dag\dag}$    & 64.75$^{\dag\dag}$  & 26.50$^{\dag\dag}$        & 37\%                  & 44.73$^{\dag\dag}$     & 60.75$^{\dag\dag}$ & 25.50$^{\dag\dag}$ \\
+ Collaborating                     & 93\%                  & 57.71                 & 73.25               & 21.25                     & 99\%                  & 61.10                  & 76.25     & 19.00           \\ \hline
\end{tabular}%

\caption{
Comparison of mediation outcomes across different behavioral strategies, using DeepSeek-V3-0324 as the backbone LLM.  
\textit{NUM = 1} indicates that one party in the case is randomly replaced with the specific behavioral strategy;  
\textit{NUM = ALL} means all parties are replaced.  
${\dag}$ and ${\dag\dag}$ indicate statistically significant differences from the \textit{Compromising} mode under the chi-squared test~\cite{greenwood1996guide} with $p < 0.05$ and $p < 0.01$, respectively.
}
\label{tab:conflict-modes}
\end{table*}

\begin{table}[t]
\centering
\setlength{\tabcolsep}{1mm} 
\begin{tabular}{lcccc}
\hline
\textbf{Setting}      & \textbf{$SR\uparrow$} & \textbf{$Sat\uparrow$} & \textbf{$Con\uparrow$} & \textbf{$LR\downarrow$} \\ \hline
Default Setting              & 82\%                  & 54.77                 & 70.96              & 24.00                     \\ \hline
+ Information Conflict  & 53\%                  & 37.55                 & 51.00              & 45.00                     \\
+ Resource Conflict     & 47\%                  & 37.23                 & 49.50              & 51.00                     \\
+ Behavioral Misconduct & 36\%                  & 31.18$^{\dag\dag}$    & 45.25              & 53.00$^{\dag}$            \\
+ Legal Issue           & 61\%                  & 39.21                 & 57.00$^{\dag}$     & 36.00$^{\dag\dag}$        \\
+ External Factors      & 68\%                  & 43.60$^{\dag\dag}$    & 56.75$^{\dag}$     & 39.75                     \\ \hline
\end{tabular}%

\caption{
Comparison of mediation outcomes across different dispute causes, using DeepSeek-V3-0324 as the backbone LLM. 
${\dag}$ and ${\dag\dag}$ indicate statistically significant differences from the \textit{Information Conflict} setting under the chi-squared test, with $p < 0.05$ and $p < 0.01$, respectively.
}
\label{tab:dispute-causes}
\end{table}

\subsection{Evaluator}
Our evaluation module includes two components: an outcome evaluator for assessing mediation results and a solution evaluator for measuring solution quality.

\noindent\textbf{Outcome Evaluator.}  
To assess mediation outcomes, we adopt four complementary evaluation metrics: Success Rate, Satisfaction, Consensus, and Litigation Risk, each reflecting a distinct evaluative dimension. These metrics fall into two categories based on their perspective: Success Rate and Satisfaction are \textit{subjective metrics}, while Consensus and Litigation Risk are \textit{objective metrics}. (1) \textbf{Success Rate} ($SR$) captures explicit acceptance: at the end of each mediation, all disputants are required to indicate their stance on the final solution by selecting from \{\texttt{Accept}, \texttt{Uncertain}, \texttt{Reject}\}. A mediation is considered successful only if all parties choose \texttt{Accept}. Let $N_{\text{succ}}$ and $N_{\text{tot}}$ denote the number of successful and total mediation cases, respectively. We define the success rate as:
\begin{equation}
SR = \frac{N_{\text{succ}}}{N_{\text{tot}}} \times 100\%.
\end{equation}
(2) \textbf{Satisfaction} ($Sat$) reflects each party's subjective evaluation of the outcome. After the mediation concludes, each disputant rates their level of satisfaction using a five-point Likert scale~\cite{jebb2021review}, with labels \{\texttt{Very Low}, \texttt{Low}, \texttt{Medium}, \texttt{High}, \texttt{Very High}\} mapped to scores \{0, 1, 2, 3, 4\}. The final satisfaction score is computed as a normalized weighted average:
\begin{equation}
Sat = \frac{1}{4} \times \left( \frac{\sum_{i=1}^{5} c_i \cdot (i-1)}{\sum_{i=1}^{5} c_i} \right) \times 100,
\end{equation}
where $c_i$ denotes the number of responses at level $i$. Unlike Success Rate, Satisfaction captures a more nuanced and deeper evaluation, as even accepted cases may score low in satisfaction if parties feel emotionally unfulfilled. (3) \textbf{Consensus ($Con$) and Litigation Risk ($LR$)} are evaluated from a third-party perspective, based on the full trajectory of the mediation. Consensus measures the extent to which the disputants converge on key issues, while Litigation Risk estimates the likelihood that the dispute may escalate into formal legal proceedings. Both $Con$ and $LR$ are rated on the same five-point Likert scale and computed using the same normalization formula as $Sat$. Our evaluation follows the LLM-as-a-judge paradigm~\cite{li2024llms}, a widely used approach for assessing generative model outputs. To further enhance the quality and transparency of evaluation, we follow the research ~\cite{liu2023alignbench,wang2024user} by incorporating chain-of-thought reasoning into the prompts, encouraging the LLM to reason step-by-step before generating a final judgment. Prompts are in Appendix Section 9.

\noindent\textbf{Solution Evaluator.} This component focuses on assessing the reasonableness and legal soundness of the mediator's proposed solutions. We evaluate two core elements:  
(1) the points of contention, i.e., the alignment between the human-annotated contention points ($DPoints$) and the mediator-generated points ($MDPoints$);  
(2) the legal bases, i.e., the set of legal bases provided by human experts ($DBases$) versus those proposed by the mediator ($MDBases$). For the comparison of contention points, we apply ROUGE-L and BERTScore~\cite{zhang2019bertscore}. For the evaluation of legal bases coverage, we compute Recall.


\section{Experimental Setup}
To test \textit{AgentMediation}, we empirically investigate (1) the reliability of its generated mediation processes, (2) its ability to support analysis of key mediation variables, and (3) the consistency between our automatic evaluation metrics and human judgments. For human evaluations and annotations, we recruit legal experts from prominent law schools. All participants have passed the National Unified Legal Professional Qualification Examination. To ensure fair and motivating compensation, we offer an average hourly wage of \$8.25, well above locally mandated minimum wage. To enable controlled experiments, we define a default setting for \textit{AgentMediation}. In this setting, we fix the number of mediators to one and do not assign any predefined behavioral strategies to either disputants or the mediator. Additionally, the mediator has no access to external knowledge sources, and the bargaining stage is limited to 5 dialogue rounds. To ensure reproducibility, the temperature of LLM is set to 0.

\section{Experimental Results}


\subsection{Reliability of Our Simulation (RQ1)}


To evaluate the reliability of our simulation, we conduct two human-model comparison experiments under the default setting: utterance-level and case-level.

\textbf{Utterance-Level.}
We formulate the task as role-based utterance generation, where both LLM agents and legal experts generate the next utterance for a specified role (mediator or disputant) given the same dialogue context. To ensure diversity in testing scenarios, we uniformly sample 50 dialogue cases (3–15 turns; 28 mediator and 22 disputant roles). For each case, three legal experts independently generate role-specific responses as equally valid references, and the agent responds under the same conditions.
Model outputs are evaluated using BERTScore (F1) and LLMScore (a 0–1 semantic similarity score assessed by GPT-4o). We report: (1) the average similarity with each expert, and (2) the average of the highest score per case as an upper bound of human alignment.
As shown in Table~\ref{realism-exp}, strong LLMs such as DeepSeek-V3-0324 and GPT-4o achieve BERTScore (LLMScore) values above 0.71 (0.79). As an upper bound for human-human similarity, we take the highest similarity score among the three expert pairs in each case and average over all cases, yielding 0.69 (BERTScore) and 0.76 (LLMScore). These results suggest that our agents exhibit strong semantic alignment with expert responses. We select DeepSeek-V3-0324 as the default backbone for its balance of quality, efficiency, and open-source availability.

\textbf{Case-Level.}
To further assess simulation reliability at the case level, we employ legal experts to reconstruct the detailed mediation processes of 20 real disputes. Using the same case descriptions, we use \textit{AgentMediation} (based on DeepSeek-V3-0324) to generate simulated mediation processes for these 20 disputes. Legal experts then conduct pairwise preference evaluations between the human reconstructed and system generated versions across 18 fine-grained aspects. Each case is independently scored by three experts, and final judgments are determined via majority vote. As shown in Table~\ref{realism-exp-2}, our simulations are indistinguishable from real mediation processes across most evaluation criteria, and even exceed them in certain aspects. These results provide further evidence of our simulation’s reliability.

\subsection{The Effect of Different Factors (RQ2)}
Having validated the reliability of \textit{AgentMediation}, we analyze how key factors influence outcomes, using over 14,000 simulated processes generated from 330 real disputes.

\subsubsection{The Effect of Behavioral Strategies}
\label{section:modes}
We investigate how disputants' behavioral strategies affect mediation outcomes. Based on the TKI Conflict Mode, we examine five typical strategies: Compromising, Competing, Accommodating, Avoiding, and Collaborating. We evaluate two configurations: (1) one randomly selected party is replaced with the specific behavioral strategy, and (2) all parties are replaced. As shown in Table~\ref{tab:conflict-modes}, we can find the following findings:

\textbf{Overall Trends Align with Theoretical Expectations.}
Under the \textit{NUM = 1} setting, the five behavioral strategies already exhibit typical patterns. (1) Competing produces the poorest results: low success (29\%), low satisfaction (43.20), low consensus (53.25), and the highest litigation risk (40.50), reflecting its escalation‑prone nature. (2) Collaborating shows the opposite extreme, achieving high success (93\%), the best satisfaction (57.71), strong consensus (73.25), and the lowest risk among active modes (21.25), consistent with its integrative, win–win orientation. (3) Compromising yields moderate results across all metrics: success 95\%, satisfaction 55.72, consensus 72.50, and risk 22.50. (4) Avoiding lowers visible risk (26.50) but also depresses satisfaction (49.63) and consensus (64.75), showing that backing away reduces open conflict yet leaves problems unsolved. (5) Accommodating achieves high consensus (71.50) with low risk (21.00), though satisfaction remains modest (54.21), indicating that harmony may come at the cost of personal needs. These strategy‑specific tendencies persist under the \textit{NUM = ALL} setting.

\textbf{Evidence of Group Polarization~\cite{moscovici1969group} under Full Replacement.}  
Comparing \textit{NUM = 1} and \textit{NUM = ALL} reveals a clear group polarization effect: when all parties adopt the same behavioral strategy, the collective dynamics become more extreme.  
In the Competing setting, full replacement significantly reduces performance, with success rate dropping from 29\% to 11\% and litigation risk rising from 40.50 to 64.00.  
This suggests that mutual aggressiveness amplifies conflict escalation and breakdown.  
In contrast, Collaborating under \textit{NUM = ALL} leads to gains across all metrics compared to partial replacement (e.g., consensus increased from 73.25 to 76.25), reflecting a synergistic dynamic when cooperation is mutual.  
These findings support the theoretical expectation that interaction symmetry can either reinforce cooperation or escalate conflict, depending on the nature of the strategy.

\textbf{Surface-level Consensus~\cite{kelman1958compliance,habermas1985theory} in the Accommodating Strategy.}  
The discrepancy between high consensus (75.50) and relatively low satisfaction (53.35) in the Accommodating suggests the presence of a Surface-level Consensus, where disputants agree outwardly but remain internally dissatisfied.  
This occurs when individuals suppress their own needs to preserve harmony, leading to unspoken tensions or unmet expectations.  
Compared to the Avoiding strategy, which also yielded low satisfaction but lower consensus, the Accommodating strategy appears more effective in reaching surface-level agreements, but potentially at the cost of long-term resolution quality.  
This aligns with prior research cautioning against over-reliance on accommodative behaviors, especially when structural issues remain unaddressed.

\begin{table*}[t]
\small
\centering
\setlength{\tabcolsep}{1mm} 
\begin{tabular}{llcccccc}
\hline
\multirow{2}{*}{} & \multirow{2}{*}{\textbf{Different Backbone LLM}} 
& \multicolumn{2}{c}{\textbf{Points of Contention}} 
& \multicolumn{4}{c}{\textbf{Legal Bases} \textit{(Recall $\uparrow$)}} \\ 
\cmidrule(lr){3-4} \cmidrule(lr){5-8}
& & \textit{BERTScore (F1) $\uparrow$} & \textit{ROUGE-L (F1) $\uparrow$}
  & w/o EK & w/ EK (top-3) & w/ EK (top-5) & w/ EK (top-10) \\ 
\hline

\multirow{4}{*}{Open-Source} 
& Llama-3.1-70B-Instruct & 0.7647 & 0.3847 & 0.0295 & 0.1151 & 0.1392 & 0.1689 \\
& GLM-4-Flash            & 0.7994 & 0.4640 & 0.1742 & 0.2260 & 0.2461 & 0.2718 \\
& DeepSeek-V3-0324       & 0.8247 & \underline{0.5404} & \textbf{0.2999} & \textbf{0.3335} & \textbf{0.3446} & \textbf{0.3594} \\
& DeepSeek-R1            & 0.7994 & 0.4783 & \underline{0.2988} & \underline{0.3302} & 0.3386 & 0.3542 \\
\hline

\multirow{3}{*}{Closed-Source} 
& GLM-4-Plus             & \textbf{0.8354} & \textbf{0.5581} & 0.2771 & 0.3277 & 0.3395 & \underline{0.3593} \\
& GPT-4o-mini-2024-07-18           & \underline{0.8260} & 0.5288 & 0.0624 & 0.1442 & 0.1668 & 0.1939 \\
& Qwen-Plus              & 0.8231 & 0.5176 & 0.2917 & 0.3301 & \underline{0.3411} & 0.3578 \\
\hline
\end{tabular}%

\caption{Performance comparison across different backbone LLMs under different evaluation metrics. EK denotes access to External Knowledge. The top-$k$ represents different retrieval settings, where $k$ represents the number of legal bases retrieved.
}
\label{tab:my-table}

\end{table*}

\subsubsection{The Effect of Dispute Causes}
\label{section:causes}

In this section, we investigate how different dispute causes affect mediation outcomes. While each case in our dataset is associated with a fixed legal dispute type (e.g., housing contract), the underlying causes can vary considerably in practice. To analyze their specific impact, we construct controlled variants of each case by injecting or amplifying a particular dispute cause within the Dispute Brief ($DB$). We consider five representative dispute causes, defined by the two-level taxonomy introduced earlier: Information Conflict, Resource Conflict, Behavioral Misconduct, Legal Issues, and External Factors. Given a specified top-level cause, we first prompt DeepSeek-V3-0324 to select the most contextually appropriate sub-cause based on the original $DB$. Then, conditioned on the selected sub-cause, the LLM generate targeted modifications that reinforce the presence of the cause in the $DB$. As Table~\ref{tab:dispute-causes} shows, our main findings include:

\textbf{Moral Emotions Theory}~\cite{haidt2003moral} suggests that perceiving others' misconduct, such as deception or intentional rule-breaking, can elicit moral emotions like anger, contempt, and disgust. These emotions often lead to punitive or distancing responses, making compromise more difficult in mediation. This pattern aligns with our empirical observations: cases involving Behavioral Misconduct exhibit the highest litigation risk (53.00), the lowest satisfaction (31.18), and the lowest resolution success rate (36\%).
 
\textbf{Realistic Conflict Theory}~\cite{Sherif1961RobbersCave} argues that disputes perceived as zero-sum competition over limited resources lead to increased hostility.  
Our Resource Conflict cases support this view, with elevated risk (51.00), low satisfaction (37.23), and a reduced success rate of 47\%.

\textbf{The Common In‑Group Identity Model}~\cite{gaertner1993common} states that facing a common external threat can blur group boundaries and encourage cooperation.  Consistent with this view, disputes driven by External Factors (e.g., policy shocks or natural disasters) and Legal Issues achieve the highest consensus scores (56.75 and 57.00) and elevated success rates (68\% and 61\%).  Litigation risk is also lower for Legal Issues (36.00), suggesting that when parties frame the problem as procedural rather than personal, they are more willing to compromise and avoid escalation.

\subsubsection{The Effect of Mediator Expertise}
\label{section:LLM}

This section examines how mediator expertise influences mediation outcomes. As shown in Table~\ref{tab:my-table}, DeepSeek-V3-0324 outperforms all open-source LLMs and rivals closed-source LLMs, achieving strong results in contention identification (BERTScore 0.8247, ROUGE-L 0.5404) and the highest recall in legal citation. These results make it a highly competitive backbone LLM in our \textit{AgentMediation}. Moreover, all LLMs benefit from External Knowledge. For instance, DeepSeek-V3-0324’s recall improves to 0.3594 with top-10 retrieval, showing that retrieval-augmented generation (RAG) can effectively enhance domain-specific performance.

Overall, the above experiments addressing RQ2 show that \textit{AgentMediation} can effectively explore the impact of key mediation variables and uncover phenomena consistent with established social theories, highlighting its ability to capture legal-social dynamics and its practical value for advancing legal research and intelligent mediation.

\subsection{Reliability of Our LLM-as-a-Judge (RQ3)}
To assess the reliability of our LLM-as-a-judge, we conduct a human-model agreement study on the three key outcome metrics: Satisfaction, Consensus, and Litigation Risk. For each, we randomly sample 50 representative mediation cases. Nine legal experts are recruited, with three assigned per metric to independently score all cases. We calculate agreement between model predictions and the majority vote of the three expert annotations per case. Using DeepSeek-V3-0324 as the judge, we observe Cohen’s Kappa scores of 0.358 for Satisfaction, 0.519 for Consensus, and 0.672 for Litigation Risk. These results suggest that while Satisfaction remains the most subjective metric (even inter-annotator agreement is low, at 0.344), both the LLM-as-a-judge's predictions for Consensus and Litigation Risk show strong correlation with human annotations, providing empirical support for leveraging LLM-as-a-judge in large-scale mediation simulations. Additional details are in Appendix Section 3.

\subsection{Ablation Studies and Discussions}

Beyond the main experiments, we conducted a broad range of in-depth follow-up studies. Due to space limitations, we summarize the key experiments and conclusions here; detailed results are provided in the appendix. (1) We performed ablations under the default setting to assess the contribution of individual mediation stages (Section 4) and examined the impact of varying bargaining rounds (Section 5). Results show that removing any single stage leads to a noticeable decline in mediation performance, while increasing the number of bargaining turns yields only marginal gains. (2) To verify our findings are not model-specific, we replicated key experiments using GLM-4-Plus and observed consistent trends, confirming the robustness of our conclusions across different backbone LLMs (Section 6). (3) We further extended our analysis beyond single-variable ablations by exploring multi-variable interactions (Section 7) and enabling dynamic strategy adaptation (Section 8). These studies produced interpretable results, further validating the flexibility of our framework. (4) Lastly, \textit{AgentMediation} is designed for cross-cultural adaptability. By incorporating the widely recognized HDR framework and TKI mode, it could be applied broadly across legal systems and cultural contexts.

\section{Conclusion}

We present \textit{AgentMediation}, the first LLM-based agent framework for simulating dispute mediation. Grounded in real-world disputes, it produces realistic mediation processes and serves as an extensible platform for controlled studies on key factors. Our simulations align with established sociological theories and offer insights into mediation dynamics. Looking forward, our framework provides a foundation for developing intelligent mediation systems.

\bibliography{aaai2026}

\section{Appendix}
\subsection{1. Dataset Details}
\label{section:casedetail}
\begin{table*}[t]
\centering
\resizebox{\textwidth}{!}{%
\begin{tabular}{p{\textwidth}}
\hline
\textbf{Title}:\\ Mediation of a Housing Sale Contract Dispute Between Sun and Gong \\ \hline

\textbf{Keywords}:\\ civil dispute, housing contract, government-managed housing, refund amount... \\ \hline

\textbf{Brief}:\\  
In August 2009, Sun and Gong signed a housing transfer contract through an agency. Sun agreed to transfer the usage rights of a government-managed public housing unit to Gong for 6,700 RMB per square meter. Although the payment and physical handover were completed, transfer procedures were delayed due to emerging disputes. In 2023, the local government halted the transfer of such public housing rights, rendering the contract unenforceable. Sun filed a lawsuit in June 2023 seeking to reclaim use of the property. \\ \hline

\textbf{Method}: \\ 
With both parties’ consent, the court initiated a mediation process through a digital platform linked with the National Development and Reform Commission's Price Certification Center, with participation from the Suzhou Price Certification Bureau. Authorities confirmed that the property could no longer be transferred due to policy restrictions. The core issue shifted to determining a fair refund amount.  
To address this, mediators conducted on-site inspections and market analysis of similar privately-owned properties, considering location, structure, layout, surrounding environment, rent levels, and relocation policies. Based on this, a reference price of 20,000 RMB per square meter was proposed and explained professionally. \\ \hline

\textbf{Outcome}: \\ 
Both parties accepted the proposed refund terms. Sun withdrew the lawsuit. \\ \hline

\textbf{Bases}:  \\
Civil Code of the People’s Republic of China: Articles 533, 563, and 566.  \\
Opinion on Promoting Price Dispute Mediation (Fa Fa [2019] No. 32). \\ \hline
\end{tabular}%
}
\caption{A raw case input example before data preprocessing.}
\label{tab:raw_case}
\end{table*}
\begin{table*}[t]
\centering
\resizebox{\textwidth}{!}{%
\begin{tabular}{p{\textwidth}}
\hline
\textbf{Title}: \\
Mediation Case Between Sun and Gong Over Housing Sale Contract \\ \hline

\textbf{Dispute Type ($DT$)}: \\
housing contract \\ \hline

\textbf{Dispute Brief ($DB$)}:  \\
In August 2009, Sun and Gong signed a housing transfer contract via a real estate agency. Sun voluntarily agreed to transfer the usage rights of a government-managed public housing unit to Gong at a rate of 6,700 RMB per square meter. After payment and property handover were completed, a dispute arose, and the ownership transfer was delayed. In 2023, the local government halted transactions of such housing rights, making resolution difficult. In June 2023, Sun filed a lawsuit to reclaim actual usage of the property. \\ \hline

\textbf{Dispute Facts ($DF$)}:  \\
1. Housing Transfer Contract — In August 2009, Sun and Gong signed a housing transfer agreement. Sun agreed to transfer the usage rights of a government-managed public housing unit to Gong for 6,700 RMB/m².\\
2. Policy Document — In 2023, local government policies terminated the transferability of government-managed housing rights.\\
3. On-site Inspection — The mediator conducted field inspections, evaluating the property's location, structure, layout, and surroundings. Market prices of comparable private properties were also investigated.\\
4. Market Research — The mediator incorporated rental data and relocation policy factors to estimate a reasonable refund amount.
 \\ \hline

\textbf{Dispute Parties ($DP$)}:  \\
Sun, Gong \\ \hline

\textbf{The Points of Contention ($DPoints$)}: \\ 
The core dispute lies in whether the agreed refund amount complies with the contract and relevant legal provisions, given that the contract could not be fulfilled due to policy changes. \\ \hline

\textbf{Dispute Bases ($DBases$)}:  \\
1. Civil Code of the People’s Republic of China, Article 533 \\
2. Civil Code of the People’s Republic of China, Article 563 \\
3. Civil Code of the People’s Republic of China, Article 566 \\
4. Opinion on Deepening Price Dispute Mediation (Fa Fa [2019] No. 32)
\\ \hline
\end{tabular}%
}
\caption{A structured case example after data preprocessing}
\label{tab:structured_case}
\end{table*}

\begin{figure}[t]
    \centering
    \includegraphics[width=\linewidth]{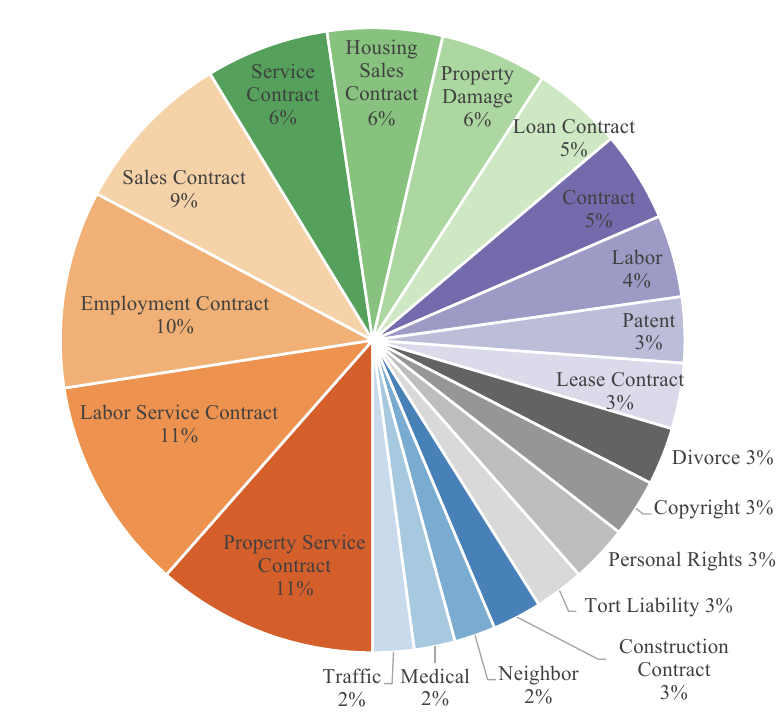}
    \caption{The distribution of dispute types.}
    \label{fig:dispute-types}
\end{figure}

To provide a concrete understanding of our data preprocessing pipeline, this section presents a full case example drawn from the \textit{Dispute Resolution Case Database}. We include both (1) the original case (Table~\ref{tab:raw_case}), and (2) its structured version after preprocessing (Table~\ref{tab:structured_case}). In addition, Figure~\ref{fig:dispute-types} illustrates the distribution of the top 20 dispute case types included in our dataset.

\subsection{2. Dispute Causes Taxonomy Details}
\label{appendix: dispute-causes}

\begin{figure*}[t]
    \centering
    \includegraphics[width=\textwidth]{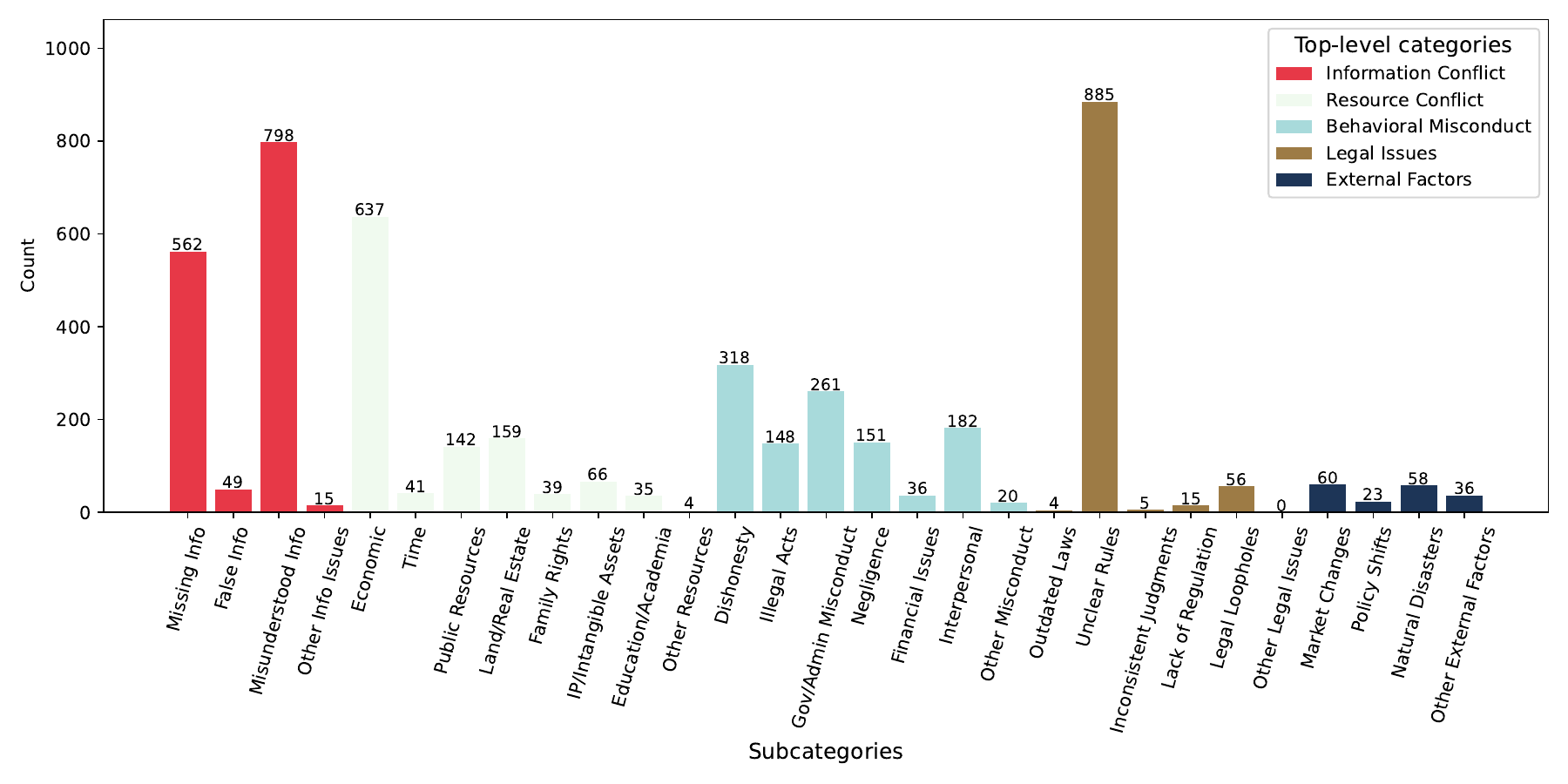}
    \caption{The distribution of dispute causes across 1,080 civil court cases. Bars with the same color represent subcategories under the same top-level cause type.}
    \label{fig:dispute-causes}
\end{figure*}

In this section, we detail the construction of our \textbf{Dispute Causes Taxonomy}. We begin by designing an initial taxonomy grounded in prior legal scholarship~\cite{vilendrer2023conflict}. Building on this foundation, we employ GPT-4o to infer the underlying causes of 1,080 real-world civil court cases based on their factual descriptions. When cases fall outside the scope of the existing taxonomy, we extend it dynamically to incorporate newly emerging cause types. To ensure both coverage and interpretability, the resulting taxonomy is reviewed and refined by legal experts. The finalized taxonomy is illustrated in Figure~\ref{fig:dispute-causes}, where bars of the same color denote categories under the same top-level cause type. Our taxonomy consists of five major categories: \textbf{Information Conflict, Resource Conflict, Behavioral Misconduct, Legal Issues, and External Factors}. Each category contains multiple subcategories, along with an \textbf{Other} class to preserve completeness. The figure further presents the distribution of causes derived from the 1,080 civil court cases. Two key insights emerge:

(1) The \textbf{Other} subcategories in each top-level group appear infrequently, suggesting that the taxonomy captures the vast majority of common causes and achieves high coverage.

(2) Many cases involve multiple interrelated causes. For example, \textbf{Misunderstood Info} and \textbf{Unclear Rules} are among the most frequently co-occurring subcategories, indicating that ambiguity in legal frameworks often accompanies miscommunication. This reflects the inherently multifactorial and intertwined nature of legal disputes, which calls for a multi-causal modeling approach.

Notably, our taxonomy is designed to be extensible. As new patterns of disputes emerge, the framework can be readily expanded to accommodate evolving legal scenarios, supporting long-term applicability in real-world legal NLP systems.

\subsection{3. Details of the Reliability Evaluation of Our LLM-as-a-Judge}

To evaluate the reliability of our LLM-as-a-judge, we conducted a human-model agreement study for the three key outcome metrics in our framework: Satisfaction, Consensus, and Litigation Risk. For each metric, we randomly sampled 50 representative mediation cases.  We recruited nine legal experts, each of whom had passed the National Unified Legal Professional Qualification Examination, and assigned three different experts per metric to independently score all 50 cases. For evaluation, we computed:

\noindent(1) Human–human agreement (“Among Human”): the average pairwise agreement across the three annotators per metric.

\noindent(2) Model–human agreement: the agreement between model predictions and the majority vote of the three human labels per case.

\noindent(3) Model–model agreement: the alignment between different LLMs on the same evaluation task.

We used four standard metrics: Pearson correlation (P), Spearman’s rank correlation (S), Kendall’s Tau (T), and Cohen’s Kappa (K), covering both continuous and categorical perspectives.

Tables \ref{Satisfaction}, \ref{Consensus} and \ref{Litigation Risk} present the results, showing that Satisfaction is the most subjective and variable dimension. The inter-annotator agreement among human experts on Satisfaction was moderate (Cohen’s Kappa = 0.344), while the agreement between LLMs and humans ranged from 0.229 to 0.358. Among the evaluated models, DeepSeek-V3-0324 achieved the highest alignment with human ratings (Kappa = 0.358), followed by GPT-4o (0.302) and GLM-4-Plus (0.229). These findings highlight the inherent complexity and subjectivity of measuring emotional or psychological outcomes, even among legal professionals. In contrast, both Consensus and Litigation Risk exhibited substantially higher and more stable agreement. The inter-human agreement reached Kappa scores of 0.529 and 0.613 for Consensus and Litigation Risk, respectively. Similar levels of agreement were observed between model predictions and human judgments. Specifically, DeepSeek-V3-0324 attained a Kappa of 0.519 (Consensus) and 0.672 (Litigation Risk); GPT-4o reached 0.533 and 0.704; and GLM-4-Plus achieved 0.541 and 0.681, respectively. Agreement between model pairs was also high, with Kappa values exceeding 0.67 across all comparisons, often surpassing 0.80.

These results suggest that while Satisfaction remains a challenging metric due to its subjective nature, Consensus and Litigation Risk can be judged consistently and reliably by both human experts and LLMs. This provides empirical support for the use of LLMs as evaluators in large-scale mediation simulations. In particular, DeepSeek-V3-0324 demonstrated high alignment with human labels on both dimensions, reinforcing its suitability as the backbone LLM for our framework.

\begin{table}[t]
\resizebox{\columnwidth}{!}{%
\begin{tabular}{lcccc}
\hline
\textbf{}                      & \textbf{P} & \textbf{S} & \textbf{T} & \textbf{K} \\ \hline
Among Human                    & 0.356      & 0.354      & 0.326      & 0.344      \\
DeepSeek-V3-0324 vs Human      & 0.376      & 0.358      & 0.330      & 0.358      \\
GPT-4o vs Human                & 0.481      & 0.450      & 0.422      & 0.302      \\
GLM-4-Plus vs Human            & 0.303      & 0.290      & 0.273      & 0.229      \\ \hline
DeepSeek-V3-0324 vs GPT-4o     & 0.594      & 0.577      & 0.557      & 0.409      \\
DeepSeek-V3-0324 vs GLM-4-Plus & 0.662      & 0.670      & 0.647      & 0.557      \\
GPT-4o vs GLM-4-Plus           & 0.534      & 0.534      & 0.534      & 0.504      \\ \hline
\end{tabular}%
}
\caption{Agreement Analysis Between Humans and LLMs on \textbf{Satisfaction}.}
\label{Satisfaction}
\end{table}

\begin{table}[t]
\resizebox{\columnwidth}{!}{%
\begin{tabular}{lcccc}
\hline
\textbf{}                      & \textbf{P} & \textbf{S} & \textbf{T} & \textbf{K} \\ \hline
Among Human                    & 0.604      & 0.610      & 0.554      & 0.529      \\
DeepSeek-V3-0324 vs Human      & 0.668      & 0.649      & 0.621      & 0.519      \\
GPT-4o vs Human                & 0.648      & 0.651      & 0.617      & 0.533      \\
GLM-4-Plus vs Human            & 0.651      & 0.653      & 0.618      & 0.541      \\
DeepSeek-V3-0324 vs GPT-4o     & 0.863      & 0.883      & 0.869      & 0.853      \\
DeepSeek-V3-0324 vs GLM-4-Plus & 0.831      & 0.849      & 0.836      & 0.820      \\
GPT-4o vs GLM-4-Plus           & 0.906      & 0.928      & 0.913      & 0.905      \\ \hline
\end{tabular}%
}
\caption{Agreement Analysis Between Humans and LLMs on \textbf{Consensus}.}
\label{Consensus}
\end{table}

\begin{table}[t]
\resizebox{\columnwidth}{!}{%
\begin{tabular}{lcccc}
\hline
\textbf{}                      & \textbf{P} & \textbf{S} & \textbf{T} & \textbf{K} \\ \hline
Among Human                    & 0.619      & 0.654      & 0.633      & 0.613      \\
DeepSeek-V3-0324 vs Human      & 0.678      & 0.739      & 0.717      & 0.672      \\
GPT-4o vs Human                & 0.719      & 0.751      & 0.732      & 0.704      \\
GLM-4-Plus vs Human            & 0.705      & 0.744      & 0.716      & 0.681      \\
DeepSeek-V3-0324 vs GPT-4o     & 0.699      & 0.716      & 0.696      & 0.679      \\
DeepSeek-V3-0324 vs GLM-4-Plus & 0.789      & 0.791      & 0.766      & 0.765      \\
GPT-4o vs GLM-4-Plus           & 0.733      & 0.704      & 0.690      & 0.729      \\ \hline
\end{tabular}%
}
\caption{Agreement Analysis Between Humans and LLMs on \textbf{Litigation Risk}.}
\label{Litigation Risk}
\end{table}

\subsection{4. Ablation Study of Mediation Stages}
\begin{table}[t]
\centering
\setlength{\tabcolsep}{1mm} 
\small
\begin{tabular}{lcccc}
\toprule
\textbf{Setting} & $SR$ $\uparrow$ & $Sat$ $\uparrow$ & $Con$ $\uparrow$ & $LR$ $\downarrow$ \\
\midrule
Full five-stage      & 92\% & 56.82 & 73.75 & 26.75 \\
w/o Statement              & 87\% & 55.96 & 73.25 & 26.75 \\
w/o Option Generation      & 84\% & 55.11 & 70.50 & 27.75 \\
w/o Closure                & 61\% & 53.85$^{\dag\dag}$ & 68.50$^{\dag\dag}$ & 30.75$^{\dag}$ \\
\bottomrule
\end{tabular}
\caption{Ablation results of different mediation stages under the default setting, based on a subset of 100 sampled cases.}
\label{ablation}
\end{table}

We conducted ablation studies under the default setting to examine the impact of individual stages within our mediation pipeline, based on a subset of 100 sampled cases. As shown in Table~\ref{ablation}, removing any single stage leads to a decline in mediation performance, highlighting that each component of the five-stage pipeline plays a distinct and complementary role in enabling coherent and realistic mediation simulations.

\begin{table*}[t!]
\renewcommand{\arraystretch}{1.5}
\centering
\resizebox{\textwidth}{!}{%
\begin{tabular}{l|cccc|cccc}
\hline
\multirow{2}{*}{\textbf{Setting}} & \multicolumn{4}{c|}{\textbf{NUM = 1}}                     & \multicolumn{4}{c}{\textbf{NUM = ALL}}                      \\ \cline{2-9}
                                  & \textbf{\begin{tabular}[c]{@{}c@{}}Success Rate\\ ($SR\uparrow$)\end{tabular}} 
                                  & \textbf{\begin{tabular}[c]{@{}c@{}}Satisfaction\\ ($Sat\uparrow$)\end{tabular}} 
                                  & \textbf{\begin{tabular}[c]{@{}c@{}}Consensus\\ ($Con\uparrow$)\end{tabular}} 
                                  & \textbf{\begin{tabular}[c]{@{}c@{}}Litigation Risk\\ ($LR\downarrow$)\end{tabular}} 
                                  & \textbf{\begin{tabular}[c]{@{}c@{}}Success Rate\\ ($SR\uparrow$)\end{tabular}} 
                                  & \textbf{\begin{tabular}[c]{@{}c@{}}Satisfaction\\ ($Sat\uparrow$)\end{tabular}} 
                                  & \textbf{\begin{tabular}[c]{@{}c@{}}Consensus\\ ($Con\uparrow$)\end{tabular}} 
                                  & \textbf{\begin{tabular}[c]{@{}c@{}}Litigation Risk\\ ($LR\downarrow$)\end{tabular}} \\ \hline
Default Setting                          & 71\% & 66.87 & 68.00 & 27.00 & 71\% & 66.87 & 68.00 & 27.00     \\ \hline
+ Compromising                      & 80\% & 68.66 & 71.50 & 27.50 & 87\% & 70.03  & 73.75  & 25.75  \\
+ Competing                         & 42\% & 58.51$^{\dag\dag}$ & 56.25$^{\dag\dag}$ & 45.50$^{\dag\dag}$ & 18\% & 53.08$^{\dag\dag}$ & 46.50$^{\dag\dag}$ & 66.00$^{\dag\dag}$ \\
+ Accommodating                     & 79\% & 68.60 & 70.50 & 26.75 & 75\% & 68.01$^{\dag}$ & 73.23$^{\dag}$ & 25.25 \\
+ Avoiding                          & 69\% & 66.89 & 68.50 & 28.00 & 45\% & 61.90$^{\dag\dag}$ & 61.75$^{\dag\dag}$ & 28.50$^{\dag}$ \\
+ Collaborating                     & 78\% & 69.34 & 70.50 & 27.25 & 82\% & 71.18  & 74.25  & 25.75  \\ \hline
\end{tabular}%
}
\caption{
Comparison of mediation outcomes across different behavioral strategies, using \textbf{GLM-4-Plus} as the backbone LLM. 
\textbf{NUM = 1} indicates that one party is replaced with the specific behavioral strategy;  
\textbf{NUM = ALL} means all parties are replaced.  
${\dag}$ and ${\dag\dag}$ denote statistically significant differences from the \textit{Compromising} condition under the chi-squared test with $p < 0.05$ and $p < 0.01$, respectively.
}
\label{tab:conflict-modes-glm-4-plus}
\end{table*}

\subsection{5. Ablation Study of Bargaining Turns}
\label{section:ablation}
\begin{figure}[t]
    \centering
    \includegraphics[width=\columnwidth]{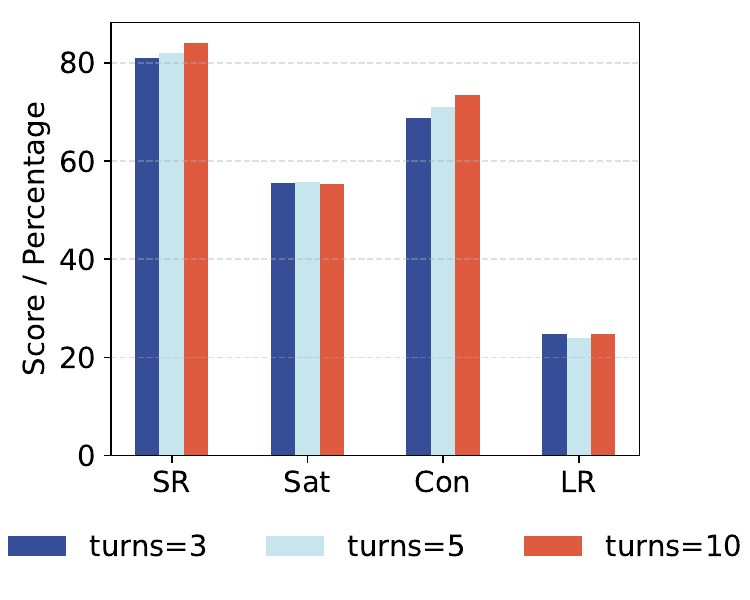}
    \caption{Effect of bargaining turns on mediation outcomes in the default setting, based on a subset of 100 sampled cases.}
    \label{fig:ablation}
\end{figure}

\textbf{Increasing the number of bargaining turns yields only marginal improvements.}
As Figure~\ref{fig:ablation} shows, we assess the impact of bargaining turns by comparing default settings with different turn counts, based on a subset of 100 sampled cases. As the number of turns increases, we observe slight improvements in success rate and consensus, suggesting that extended interaction may help parties reach better alignment. However, satisfaction and litigation risk remain largely stable, indicating that simply prolonging the dialogue does not yield significant gains in perceived quality or conflict resolution. These findings suggest that factors such as dispute causes or behavioral strategy may be more critical than interaction turns alone.

\subsection{6. Experiments Based on Different LLMs}
\label{appendix:robustness}
To evaluate the robustness of our findings across different LLMs, we replicated our main experiments using GLM-4-Plus, a closed-source LLM different from DeepSeek-V3-0324. As shown in Table~\ref{tab:conflict-modes-glm-4-plus}, the overall trends remain consistent: Compromising and Collaborating consistently achieve high success rates, satisfaction, and consensus, while Competing leads to significantly lower performance and increased litigation risk ($p < 0.01$). Moreover, the previously observed signs of Group Polarization~\cite{moscovici1969group} and Surface-level Consensus~\cite{kelman1958compliance, habermas1985theory} are also reproduced under this alternative LLM. These results align closely with those obtained using DeepSeek-V3-0324, confirming that our key observations hold robustly across different backbone LLMs.

\subsection{7. Experiments of Multi-Variable Interaction}

In our main experiments, we adopted a single-variable control design to examine how individual factors—such as behavioral strategies, dispute causes, and model choices—affect mediation outcomes. This was intended as a first step to validate the controllability and interpretability of the AgentMediation framework under clearly defined interventions.

However, in real-world settings, variables rarely appear in isolation. For example, a ``resource dispute + competitive strategy" combination may amplify conflict intensity. Different dispute causes often elicit different behavioral responses, which in turn shape dialogue dynamics and resolution likelihood. Therefore, studying only isolated variables may limit ecological validity. To address this, we conducted a multi-variable interaction experiment using our extensible framework. Specifically, we randomly sampled 100 cases to evaluate combinations of representative strategies and dispute cause scenarios, using DeepSeek-V3-0324 as the backbone LLM. The results are presented in Table~\ref{multi-varible}.

These results reveal clear nonlinear interaction effects between dispute cause and behavioral strategy. When a resource-related dispute is paired with a competitive strategy, the mediation process almost completely fails. In contrast, combining the same dispute with a collaborative strategy leads to a 90\% success rate. These findings support our AgentMediation's ability to capture complex interactions between scenario settings and agent behaviors in a realistic and interpretable way.

\begin{table*}[t]
\resizebox{\textwidth}{!}{%
\begin{tabular}{lcccc}
\hline
\textbf{Setting}                             & \textbf{Success Rate} & \textbf{Satisfaction} & \textbf{Consensus} & \textbf{Litigation Risk} \\ \hline
Default Setting                                     & 92\%                  & 56.82                 & 73.75              & 26.75                    \\
Default Setting + Competing                         & 8\%                   & 14.87                 & 3.00               & 89.50                    \\
Default Setting + Resource Conflict                 & 55\%                  & 41.19                 & 56.00              & 53.75                    \\
Default Setting + Resource Conflict + Competing     & 1\%                   & 11.07                 & 1.52               & 99.49                    \\
Default Setting + Collaborating                     & 98\%                  & 68.89                 & 82.00              & 16.25                    \\
Default Setting + Resource Conflict + Collaborating & 90\%                  & 55.38                 & 75.75              & 25.00                    \\ \hline
\end{tabular}%
}
\caption{Impact of multi-variable interaction on mediation outcomes, evaluated on a subset of 100 sampled cases using \textbf{DeepSeek-V3-0324} as the backbone LLM. All parties are assigned specified behavioral strategies.}
\label{multi-varible}
\end{table*}

\subsection{8. Experiments of Dynamic Strategies}
In our main experiments, we adopted static behavioral strategies for agents to better isolate and analyze the effects of different strategies on mediation outcomes. This design enhances the contrast between strategies and helps reveal their causal impact. It is also theoretically supported: according to the Thomas-Kilmann Conflict Model and related studies, individuals often exhibit relatively stable conflict-handling styles, especially in socially constrained roles like those seen in mediation.

However, we also noticed that real-world negotiation behavior is dynamic and adaptive. In fact, our AgentMediation framework includes a built-in reflection module that enables agents to revise their strategies based on the evolving dialogue context. This mechanism was disabled in the main experiments for clarity, but we activated it in follow-up tests to evaluate its utility. Each agent, when equipped with this module, is able to assess current dialogue status, opponent behavior, and topic focus after each turn, and then decide whether to retain or switch its strategy accordingly. (details in our code).
We conducted comparative experiments on 100 cases using two backbone LLMs (DeepSeek-V3-0324 and GLM-4-Plus), under both static and dynamic settings.

Tables~\ref{dynamic1} and~\ref{dynamic2} present the results, from which we derive the following key findings:

For the Avoiding strategy, both satisfaction and consensus scores improved substantially across the two LLMs (e.g., satisfaction rose from 29.53 to 53.05 in DeepSeek, and from 61.90 to 68.16 in GLM), suggesting that passive agents can be guided toward more constructive behavior over time through adaptive strategy adjustment.

For the Competing strategy, GLM-4 exhibited a notable increase in success rate from 18\% to 44\%, indicating that enabling dynamic strategy shifts helps mitigate negotiation breakdowns caused by rigidly confrontational behavior.

Overall, these findings demonstrate that our system can effectively model dynamic strategy adaptation, enabling the simulation of more realistic and flexible mediation scenarios.

\begin{table*}[t]
\resizebox{\textwidth}{!}{%
\begin{tabular}{lcccccccc}
\hline
\textbf{Setting} & \multicolumn{2}{c}{\textbf{Success Rate}} & \multicolumn{2}{c}{\textbf{Satisfaction}} & \multicolumn{2}{c}{\textbf{Consensus}} & \multicolumn{2}{c}{\textbf{Litigation Risk}} \\ \cline{2-9} 
                 & static              & dynamic             & static             & dynamic              & static            & dynamic            & static               & dynamic               \\ \hline
Compromising     & 99\%                & 98\%                & 67.38              & 66.18                & 79.75             & 77.50              & 20.00                & 21.00                 \\
Competing        & 8\%                 & 9\%                 & 14.87              & 19.10                & 3.00              & 7.50               & 89.50                & 86.75                 \\
Accommodating    & 95\%                & 95\%                & 47.20              & 50.84                & 75.50             & 77.00$^{\dag}$    & 18.00                & 19.25                 \\
Avoiding         & 36\%       & 85\%       & 29.53     & 53.05$^{\dag\dag}$     & 30.00    & 67.50$^{\dag\dag}$   & 41.00       & 27.25$^{\dag\dag}$      \\
Collaborating    & 98\%                & 99\%                & 68.89              & 67.74                & 82.00             & 78.54$^{\dag}$    & 16.25                & 19.44                 \\ \hline
\end{tabular}%
}
\caption{Impact of dynamic strategy adaptation on mediation outcomes, based on a subset of 100 sampled cases and using \textbf{DeepSeek-V3-0324} as the backbone LLM. All parties are assigned specified behavioral strategies. Statistically significant improvements compared to the \textbf{static setting} are marked with $\dag$ ($p<0.05$) and $\dag\dag$ ($p<0.01$) under the chi-squared test.}

\label{dynamic1}
\end{table*}

\begin{table*}[t]
\resizebox{\textwidth}{!}{%
\begin{tabular}{lcccccccc}
\hline
\textbf{Setting}   & \multicolumn{2}{c}{\textbf{Success Rate}} & \multicolumn{2}{c}{\textbf{Satisfaction}} & \multicolumn{2}{c}{\textbf{Consensus}} & \multicolumn{2}{c}{\textbf{Litigation Risk}} \\ \cline{2-9} 
                   & static              & dynamic             & static             & dynamic              & static            & dynamic            & static               & dynamic               \\ \hline
Compromising       & 87\%                & 88\%                & 70.03              & 70.56                & 73.75             & 73.50              & 25.75                & 25.50                 \\
Competing & 18\%       & 44\%       & 53.08     & 60.34$^{\dag\dag}$     & 46.50    & 56.75$^{\dag\dag}$   & 66.00       & 43.75$^{\dag\dag}$      \\
Accommodating      & 75\%                & 88\%                & 68.01              & 70.79$^{\dag\dag}$     & 73.23             & 74.25              & 25.25                & 24.75                 \\
Avoiding           & 45\%       & 64\%       & 61.90              & 68.16$^{\dag\dag}$     & 61.75             & 72.00$^{\dag\dag}$   & 28.50                & 26.25                 \\
Collaborating      & 82\%                & 83\%                & 71.18              & 71.05                & 74.25             & 74.00              & 25.75                & 25.75                 \\ \hline
\end{tabular}%
}
\caption{Impact of dynamic strategy adaptation on mediation outcomes, based on a subset of 100 sampled cases and using \textbf{GLM-4-Plus} as the backbone LLM. All parties are assigned specified behavioral strategies. Statistically significant improvements compared to the \textbf{static setting} are marked with $\dag$ ($p<0.05$) and $\dag\dag$ ($p<0.01$) under the chi-squared test.}
\label{dynamic2}
\end{table*}

\subsection{9. Evaluation Prompt Details}
\label{section:evalprompt}
To provide a clear overview of our evaluation setup, this section presents the prompt templates used to elicit different types of assessments during the mediation simulation. Specifically, we include:  
(1) the prompt for asking disputants whether they accept the final mediation outcome (Table~\ref{tab:prompt_acceptance_decision});  
(2) the Prompt for Satisfaction Rating (Table~\ref{tab:prompt_satisfaction_rating});  
(3) the Prompt for Consensus Degree Rating (Table~\ref{tab:prompt_consensus_rating}); and  
(4) the Prompt for Litigation Risk Rating (Table~\ref{tab:prompt_conflict_risk}).  

\subsection{10. Case Study}
To demonstrate the quality of our simulation, we present a representative case generated by \textit{AgentMediation}, centered on a dispute over the unauthorized use of a digital game account. As shown in Table~\ref{tab:fe_wang_full_mediation1}, \ref{tab:fe_wang_full_mediation2}, and \ref{tab:fe_wang_full_mediation3}, the dialogue displays strong coherence and emotionally grounded expressions. The mediator’s interventions are neutral and context-aware, guiding the parties toward a balanced resolution. This case highlights \textit{AgentMediation}'s potential for modeling complex legal mediation and supporting future research.

\begin{table*}[t]
\centering
\resizebox{\textwidth}{!}{%
\begin{tabular}{p{\textwidth}}
\hline
\textbf{The Prompt for asking disputants whether they accept the final mediation outcome.} \\ \hline

CASE BACKGROUND START \\
\{self.global\_prompt\} \\
CASE BACKGROUND END \\

DIALOGUE HISTORY START  \\ \{self.get\_history(history)\} \\
DIALOGUE HISTORY END \\

FINAL MEDIATION PROPOSAL START\\ \{mediation\} \\
FINAL MEDIATION PROPOSAL END \\

You are "\{self.name\}". Based on the above mediation dialogue and the final proposal, please make a genuine decision on whether you accept the mediation outcome. Do not rush to a conclusion. Instead, carefully reflect on your true concerns and interests. \\

Please output strictly in the following JSON format: \\
\{"Accept or Not": "Accept / Reject / Undecided", "Reason": "Brief explanation of your reasoning"\}
\\ \hline

\end{tabular}
}
\caption{The Prompt for asking disputants whether they accept the final mediation outcome at the end of Mediation.}
\label{tab:prompt_acceptance_decision}
\end{table*}

\begin{table*}[t]
\centering
\resizebox{\textwidth}{!}{%
\begin{tabular}{p{\textwidth}}
\hline
\textbf{The Prompt for Satisfaction Rating.} \\ \hline

CASE BACKGROUND START \\
\{self.global\_prompt\} \\
CASE BACKGROUND END \\

DIALOGUE HISTORY START \\
\{self.get\_history(history)\} \\
DIALOGUE HISTORY END \\

FINAL MEDIATION PROPOSAL START \\
\{mediation\} \\
FINAL MEDIATION PROPOSAL END \\

You are "\{self.name\}". Based on the above mediation dialogue and the final proposal, please express your \textbf{true and honest level of satisfaction}. This is not for others, not to appear cooperative — it's a reflection for yourself:

\begin{itemize}
  \item Have I been able to express my true concerns during this mediation?
  \item Were my most important needs directly addressed?
  \item To what extent does the final agreement fulfill my core expectations?
  \item Although there may still be disagreements or worries, is this outcome acceptable to me?
\end{itemize}

Please choose one of the following five levels that best represents your \textbf{genuine internal feeling}, and briefly explain your reason:

\begin{itemize}
  \item \textbf{Very Low}: I didn’t express my true concerns, or they were ignored, rejected, or suppressed. My core needs were unmet, and the process made me feel powerless, marginalized, or antagonized. I find the outcome hard to accept.
  \item \textbf{Low}: I expressed some views, but important concerns were suppressed or dismissed. My apparent cooperation was mostly to avoid conflict or not embarrass others. I outwardly accepted the result, but feel unsatisfied or regretful inside.
  \item \textbf{Medium}: I was able to express my stance, and some concerns were addressed. I felt partially heard, though some core issues were not deeply explored. The result is acceptable, but I remain hesitant or uncertain about some parts.
  \item \textbf{High}: I clearly expressed my stance, and both the other party and mediator responded positively. While I made some concessions, they were within reason. Key issues were addressed, and I basically agree with the outcome.
  \item \textbf{Very High}: I was able to fully express my thoughts, and my core needs were understood and addressed. The proposal fits my main concerns, and the process made me feel respected and supported. I feel at ease and satisfied with the result.
\end{itemize}

Please output strictly in the following JSON format: \\
\{"Satisfaction Level": "Very Low / Low / Medium / High / Very High", "Reason": "Brief explanation of your judgment"\}
\\ \hline

\end{tabular}
}
\caption{The Prompt for Satisfaction Rating at the end of Mediation.}
\label{tab:prompt_satisfaction_rating}
\end{table*}

\begin{table*}[t]
\centering
\resizebox{\textwidth}{!}{%
\begin{tabular}{p{\textwidth}}
\hline
\textbf{The Prompt for Consensus Degree Rating.} \\ \hline

\textbf{Prompt Content}:\\
Please evaluate the degree of consensus currently achieved among the disputing parties in the following mediation dialogue, and assign it a level from the following five categories: Very Low / Low / Medium / High / Very High. Provide a brief explanation of your judgment. \\

CASE BACKGROUND START \\
\{self.global\_prompt\} \\
CASE BACKGROUND END \\

DIALOGUE HISTORY START \\
\{self.get\_history(messages)\} \\
DIALOGUE HISTORY END \\

FINAL MEDIATION PROPOSAL START \\
\{mediation\} \\
FINAL MEDIATION PROPOSAL END \\

\textbf{Rating Criteria:}
\begin{itemize}
  \item \textbf{Very Low}: The parties show little meaningful expression or response. Silence, avoidance, or hostile language are common. Key issues are sidestepped. The dialogue remains superficial with no basis for negotiation and no signs of progress.
  \item \textbf{Low}: Some willingness to listen and understand is present, but interaction is mostly attitudinal. There is no deep engagement with concrete issues, and consensus intentions remain vague.
  \item \textbf{Medium}: Initial signs of consensus are emerging. Both parties show a willingness to solve problems. Positions converge on some issues or preliminary solutions begin to form. However, key matters remain unresolved and progress is still fragile.
  \item \textbf{High}: A preliminary agreement is forming on major issues. Language is concrete and interaction is constructive. A shared framework for resolution is largely in place, and substantive negotiation is underway.
  \item \textbf{Very High}: The parties have reached a clear agreement on major issues. Collaboration is active, and the direction of resolution is well-defined and executable. The process is ready for conclusion or formal agreement drafting.
\end{itemize}

Please output strictly in the following JSON format: \\
\{"Consensus Level": "Very Low / Low / Medium / High / Very High", "Reason": "Brief explanation of your judgment"\}
\\ \hline

\end{tabular}
}
\caption{The Prompt template for Consensus Degree Rating at the end of Mediation.}
\label{tab:prompt_consensus_rating}
\end{table*}

\begin{table*}[t]
\centering
\resizebox{\textwidth}{!}{%
\begin{tabular}{p{\textwidth}}
\hline
\textbf{The Prompt for Litigation Risk Rating.} \\ \hline

\textbf{Prompt Content}:\\
Please evaluate the current level of conflict escalation risk based on the following mediation dialogue, and assign it a level from the following five categories: Very Low / Low / Medium / High / Very High. Provide a brief explanation of your judgment. \\

CASE BACKGROUND START \\
\{self.global\_prompt\} \\
CASE BACKGROUND END \\

DIALOGUE HISTORY START \\
\{self.get\_history(messages)\} \\
DIALOGUE HISTORY END \\

FINAL MEDIATION PROPOSAL START \\
\{mediation\} \\
FINAL MEDIATION PROPOSAL END \\

\textbf{Rating Criteria:}
\begin{itemize}
  \item \textbf{Very Low}: Communication is smooth and emotions are calm. Views are expressed fully and rationally. Differences are minor or a preliminary consensus has been reached. Mediation is progressing substantively with little need for external intervention.
  \item \textbf{Low}: Although differences exist, parties communicate openly and use measured language. There is a clear willingness to negotiate and interactions are constructive. Mediation is sustainable and can proceed productively.
  \item \textbf{Medium}: Clear disagreements exist or mediation is slow-moving. Indicators include vague statements, frequent concessions, avoidance of substantive issues, or over-accommodation that suppresses core demands. No direct confrontation yet, but lacks meaningful progress. Close attention and timely guidance are needed.
  \item \textbf{High}: Confrontational language is frequent and communication is near breakdown. Mediation has largely stalled or been resisted. Some parties may begin favoring litigation.
  \item \textbf{Very High}: Emotions are severely out of control and communication has collapsed. Some parties explicitly express intent to resort to legal or external means.
\end{itemize}

Please output strictly in the following JSON format: \\
\{"Conflict Risk Level": "Very Low / Low / Medium / High / Very High", "Reason": "Brief explanation of your judgment"\}
\\ \hline

\end{tabular}
}
\caption{The Prompt for Litigation Risk Rating at the end of Mediation.}
\label{tab:prompt_conflict_risk}
\end{table*}

\begin{table*}[t]
\centering
\begin{tabularx}{\textwidth}{X}
\toprule
\multicolumn{1}{c}{\textbf{\large Stage 1: Preliminary}} \\
Zhang and Jian are villagers from the same village. Zhang makes a living by raising sheep, while Jian is a crop farmer. During the busy farming season, Jian, who was in poor health and unable to perform heavy labor, sought help with planting. He approached Zhang, who declined due to his need to tend to his sheep. Jian then proposed an exchange of labor, saying that he had experience herding sheep in his younger days and could watch Zhang’s flock in return. Zhang agreed to this mutual arrangement.

While herding the sheep, Jian passed by the property of a fellow villager, Jiang, where the sheep ate corn wrapped in a tarp placed in front of Jiang’s shed. From the following day through the fifth day, a total of 34 sheep died, including 22 pregnant ewes, 9 adult ewes, 2 lambs, and 1 breeding ram, with an estimated value exceeding 50,000 RMB. Between April 30 and May 5, 2024, after discovering the incident, Zhang promptly sought veterinary assistance. Although the remaining sheep were saved, two years of breeding efforts were lost.

Zhang reported the incident to the local police, who investigated and concluded that Jiang had not intentionally poisoned the feed. Zhang then filed a lawsuit, demanding compensation from both Jian and Jiang for the losses caused by the deaths of the sheep.
\\ \midrule
\multicolumn{1}{c}{\textbf{\large Stage 2: Statement}} \\
\textbf{Mediator Li Zhen:} Hello everyone. I’m Li Zhen from the Community People’s Mediation Committee, and I have been working in mediation for eight years. Today, I will remain neutral throughout the entire process to help clarify the facts and explore possible solutions. To ensure smooth communication, I would like each of you to briefly share your perspective in turn: first Zhang, then Jian, followed by Jiang. Zhang, please begin by telling us your concerns and what you are asking for.\par
\textbf{Zhang:} My name is Zhang. Sheep farming is the only source of income for my family. Jian agreed to help herd my sheep, but he failed to supervise them properly. As a result, they ate poisoned corn stored on Jiang’s property, and 34 of them died, including 22 pregnant ewes. Two years of breeding efforts were lost. I am requesting full compensation of 50,000 RMB from both Jian and Jiang. I cannot accept anything less.\par
\textbf{Jian:} I don’t think the blame should fall entirely on me. I was helping Zhang out of goodwill. How could I have known there was an issue with Jiang’s corn? To be honest, Jiang didn’t even secure it properly with the tarp, so he should also be held accountable. I can possibly come up with 10,000 RMB, but that is my absolute limit. I really cannot afford more.\par
\textbf{Jiang:} I don’t see how this is my fault. The corn was properly wrapped in a tarp and placed in front of my own shed. The sheep wandered in and ate it. That should not be my responsibility. If anyone is to blame, it should be Jian, since he was in charge of herding them. Out of goodwill, I am willing to contribute up to 1,000 RMB to help with medical expenses, but I cannot offer more than that.
\\ \midrule
\multicolumn{1}{c}{\textbf{\large Stage 3: Option Generation}} \\
Based on the available facts, this case involves shared responsibility among the three parties:

Jian, as the temporary shepherd, failed to adequately supervise the sheep;

Although Jiang had informed others that the corn had been treated with chemicals, he did not take sufficient protective measures to secure the toxic corn;

Zhang also bears some responsibility for not handling the dead sheep in a timely manner, which made it impossible to verify the cause of death.

Taking into account the financial situations and degrees of fault of each party, the suggested allocation of responsibility is as follows:

Jian: 50\% (25,000 RMB); Jiang: 30\% (15,000 RMB); Zhang: 20\% self-burden

The specific payment methods can be negotiated and made in installments (e.g., Jian pays over six months, Jiang within three months). It is also recommended that the three parties sign a written agreement specifying the following:

Zhang will withdraw the lawsuit upon receiving the first installment;

Jiang must place warning signs when storing hazardous materials in the future;

Jian should purchase temporary insurance before engaging in future labor exchanges.

This proposal aims to compensate Zhang for the majority of his loss while also taking into account Jian’s goodwill and Jiang’s practical financial limitations. If mediation fails and the case proceeds to litigation, the court may assign a higher proportion of liability (e.g., Jian up to 70\%), and enforcement of the judgment could take one to two years.\\ \bottomrule
\end{tabularx}
\caption{A Specific Simulation Example of \textit{AgentMediation} (Part 1).}
\label{tab:fe_wang_full_mediation1}
\end{table*}

\begin{table*}[t]
\centering
\begin{tabularx}{\textwidth}{X}
\toprule
\multicolumn{1}{c}{\textbf{\large Stage 4: Bargaining}}\\
\textbf{Mediator Li Zhen:}
Thank you all. You have each clearly expressed your views. The key issue now is to find a balanced solution concerning the division of responsibility and the amount of compensation. Based on the current evidence, Jian was indeed negligent in supervising the sheep, Jiang’s storage of corn posed safety risks, and Zhang suffered significant losses that warrant reasonable compensation. The current proposal suggests Jian should bear 50 percent of the responsibility, Jiang 30 percent, and Zhang the remaining 20 percent. This ratio reflects both the degree of fault and each party’s financial capacity. Let us begin with Zhang. Do you have any specific thoughts on this proposed arrangement?

\textbf{Zhang:}
Mediator Li, I’ve listened to your proposed percentages, but the losses I’ve suffered are too great. Thirty-four sheep died, including twenty-two pregnant ewes. Jian failed to keep an eye on the flock, and Jiang did not properly secure the corn. Both are responsible. I am willing to make a concession: Jian takes on 55 percent, Jiang 35 percent, and I will cover the remaining 10 percent myself. That is my bottom line. If this still cannot be accepted, I will have no choice but to pursue the matter through legal channels. I have the veterinary records to support my claim.

\textbf{Jian:}
Mediator Li, I feel that 50 percent is already too much for me. I helped out of goodwill, and I am not a professional shepherd. If Jiang had properly covered the corn, this would not have happened. He should take on a larger share of the responsibility. I can agree to 40 percent at most. Anything beyond that is simply not affordable for me, especially with elderly family members to care for.

\textbf{Jiang:}
Mediator Li, I do understand that Zhang has suffered a significant loss, but I still believe the main responsibility lies with Jian. I did wrap the corn with a tarp and placed it on my own property. No one expected the sheep to break in and eat it. However, since there is general agreement that I bear some responsibility, I am willing to go one step further and offer 2,000 RMB as humanitarian compensation. This is the maximum I can offer, and I truly cannot afford to pay more than that. 

\textbf{Mediator Li Zhen:}
Zhang, you mentioned that you are willing to bear 10 percent of the loss yourself, which already shows your sincerity. However, we also need to consider the financial capabilities of Jian and Jiang. Jian, you said you could take on up to 40 percent, but that is still 15 percent short of what Zhang is asking. Jiang, you are willing to offer 2,000 RMB, but under the proposed 30 percent share, the amount would be 15,000 RMB. That is a considerable gap. Let us first hear from Zhang. If Jian were to take on 45 percent and Jiang 25 percent, would you find that acceptable?

\textbf{Zhang:}
Mediator Li, I still feel that 45 percent and 25 percent are too low. Jian’s failure to supervise the sheep was the direct cause of the incident, and it is also a fact that Jiang did not properly cover the corn. Here is my revised position: Jian should bear 50 percent, Jiang 30 percent, and I will cover the remaining 20 percent myself. However, Jiang must agree to clearly mark any hazardous materials on his property in the future. Otherwise, I would rather take this to court to seek a fair resolution.

\textbf{Jian:}
Mediator Li, I understand that Zhang has suffered a serious loss, and I am also willing to take more responsibility. Given my financial situation, I can increase my share to 45 percent, which amounts to 22,500 RMB. This is the maximum amount I can afford at the moment. I hope Zhang can understand my difficulties. I am also committed to cooperating fully with the mediation process.

\textbf{Jiang:}
Mediator Li, I also understand that Zhang has suffered a major loss, and I am willing to take on part of the responsibility. Considering my financial situation and the group’s input, I can increase my offer to 3,000 RMB. This is a reasonable amount I can provide at this time. I hope Zhang can understand my position as well. I will do my best to cooperate with the mediation and make sure similar incidents do not happen again.

\textbf{Mediator Li Zhen:}
Based on how things are going so far, Zhang is willing to take on 20 percent of the loss, Jian has raised his share to 45 percent, and Jiang has offered 3,000 yuan. That’s already real progress compared to the initial proposal. Right now, the main sticking point is that Jiang’s compensation is still far below the 30 percent mark, which would be around 15,000 yuan. Given Jiang’s financial situation and what he can reasonably afford, I’d suggest that he consider increasing his offer a bit more, maybe to 5,000 yuan. At the same time, Zhang could think about making a small concession on the total amount, say by covering 15 percent himself. So the adjusted plan would be: Jian covers 45 percent, Jiang pays 5,000 yuan, and Zhang takes on 15 percent. The total compensation would come to about 27,500 yuan, which is a significant reduction from the original 50,000 yuan loss. Let’s now go around and hear from each of you. Do you think this revised arrangement could work?

\textbf{Zhang:}
Mediator Li, I’m okay with Jian taking 45 percent, but 5,000 yuan from Jiang is really too low. That’s not even enough to cover the vet bills. Here’s what I can accept: Jiang should pay at least 8,000 yuan, I’ll take on 15 percent of the loss, but I need Jian to pay his share over six months, and Jiang to pay his within three months. That’s my bottom line.
\\ \bottomrule
\end{tabularx}
\caption{A Specific Simulation Example of \textit{AgentMediation} (Part 2).}
\label{tab:fe_wang_full_mediation2}
\end{table*}

\begin{table*}[t]
\centering
\begin{tabularx}{\textwidth}{X}
\toprule
\multicolumn{1}{c}{\textbf{\large Stage 4: Bargaining}} \\
\textbf{Jian:}
Mediator Li, I’ve already pushed myself to accept 45 percent. But now Zhang is asking Jiang to pay 8,000, and that puts me in a tough spot. We all know Jiang’s financial situation. Even coming up with 3,000 was already hard for him. That said, I still want to resolve this as soon as possible, so I’m sticking with the 45 percent and hope Zhang can be a bit more understanding of Jiang’s circumstances.

\textbf{Jiang:}
Mediator Li, 5,000 yuan is really the most I can manage. I wouldn’t be able to save that much even in a whole year of farming. Since Jian is holding to his 45 percent, I’m willing to take a step forward and go with 5,000 yuan, but I’d need to pay it in six installments. If Zhang agrees to this amount, I promise to pay on time and I’ll put up warning signs near my shed to avoid any future issues. Actually, to show my sincerity, I’ll pay the 5,000 within three months instead of six.

\textbf{Mediator Li Zhen:}
So far, the negotiation has made good progress. Zhang is willing to cover 15 percent himself, Jian is holding at 45 percent, and Jiang has offered 5,000 yuan with a promise to pay within three months. We are very close to reaching an agreement. The remaining issue is that Zhang insists Jiang should pay 8,000 yuan, while Jiang says 5,000 is his absolute limit. Considering Jiang’s financial situation and Zhang’s actual losses, I suggest a compromise. If Jiang can raise his offer to 6,000 yuan and Zhang agrees to that, the total compensation would be around 28,500 yuan, which is already a big reduction from the original 50,000 yuan loss. Let’s go one by one. Can each of you accept this compromise?

\textbf{Zhang:}
Mediator Li, 6,000 still feels low. We’re talking about 34 sheep here. But since Jiang is willing to pay within three months, I can take one more step back. I’ll accept 7,000 from Jiang, Jian keeps his 45 percent, and I’ll cover 15 percent. But Jiang must put up a warning sign in front of his shed. That part is non-negotiable.

\textbf{Jian:}
Mediator Li, I understand Zhang’s loss is serious, and I’m willing to do more. Given my financial situation, I can raise my share to 47 percent, which is 23,500 yuan. That’s really the most I can manage right now. As for Jiang, I understand his difficulties too, but I hope he can stretch a bit more, maybe go up to 6,000. That would help bring us closer to what Zhang is asking. I’m ready to cooperate fully so we can settle this soon.

\textbf{Jiang:}
Mediator Li, I really can’t come up with 7,000. I haven’t even seen any income from this year’s crops. 6,000 is already pushing it. I’d need to borrow from relatives to make it work. How about this: I’ll pay 6,000, in four monthly installments. But I want Jian to agree to buy insurance next time he helps out. That would give me some peace of mind.

\textbf{Mediator Li Zhen:}
Everyone has shown a genuine willingness to solve the issue. Right now, the only gap left is Jiang’s compensation. The difference between 6,000 and 7,000 yuan is not that big. Considering Zhang’s actual loss and Jiang’s financial limits, I’d like to suggest a middle ground. Could Jiang agree to 6,500 yuan, paid over four months, and could Zhang accept that amount? This would bring the total compensation to around 29,000 yuan, which respects Zhang’s core concerns while taking into account Jiang’s difficulties. Let’s hear from each of you in turn: Zhang, then Jian, then Jiang. What do you think of this compromise?

\textbf{Zhang:}
Mediator Li, 6,500 is something I can live with, but Jiang has to pay it off within four months, no delays. I also accept Jian’s 47 percent, but everything needs to be put in writing. From now on, if they’re helping with my sheep, they must buy insurance. Otherwise, I can’t feel secure about letting them near my animals again.

\textbf{Jian:}
Mediator Li, taking on 47 percent is already a stretch for me, but I get that Zhang’s loss is serious. If Jiang agrees to pay 6,500, I won’t object, but I’ll need to pay my part over six months. I’ll have to pull funds together from different places. As for the insurance part, maybe we can discuss that later. Let’s get the compensation settled first.

\textbf{Jiang:}
Mediator Li, 6,500 is really hard for me. My harvest hasn’t come in yet, and my relatives can’t lend me any more. But I want to resolve this quickly, so I’m willing to go up to 6,500 and pay it off in four months. I just hope we can close this matter soon. I’ll also put up a warning sign in front of my shed to avoid any future incidents. Since Jian is already taking on 47 percent, I won’t ask him to contribute any more. I understand things are hard for him too.
\\ \midrule
\multicolumn{1}{c}{\textbf{\large Stage 5: Closure}} \\
After multiple rounds of negotiation, all parties have demonstrated a genuine willingness to resolve the dispute. Based on the available facts, relevant legal principles, and taking into account each party’s financial situation and degree of fault, the following final mediation proposal is recommended:

Allocation of Responsibility: (1) Jian shall bear 47\% of the loss, amounting to 23,500 RMB. (2) Jiang shall bear 13\%, totaling 6,500 RMB. (3) Zhang shall self-cover 40\%, which amounts to 20,000 RMB.

Payment Terms: (1) Jian will pay his share in full within six months. (2) Jiang will pay his share in full within four months.

Additional Conditions: (1) Jiang must install a warning sign in front of his shed. (2) Jian must purchase temporary insurance before performing any similar labor in the future.

The total compensation under this plan is 30,000 RMB, reflecting a significant concession from the original 50,000 RMB loss. It aims to compensate Zhang for the majority of his damages while also considering Jian’s and Jiang’s financial limitations. This proposal represents the most balanced solution reached through the mediation process and is provided for all parties’ careful consideration and final decision.
\\ \bottomrule
\end{tabularx}
\caption{A Specific Simulation Example of \textit{AgentMediation} (Part 3).}
\label{tab:fe_wang_full_mediation3}
\end{table*}


\end{document}